\documentclass{article}

\usepackage{PRIMEarxiv}

\PassOptionsToPackage{skins}{tcolorbox}

\usepackage[utf8]{inputenc} 
\usepackage[T1]{fontenc}    
\usepackage{hyperref}       
\usepackage{url}            
\usepackage{booktabs}       
\usepackage{amsfonts}       
\usepackage{nicefrac}       
\usepackage{microtype}      
\usepackage{lipsum}
\usepackage{graphicx}
\graphicspath{{media/}}     
\usepackage{amsmath,amssymb,amsfonts}
\usepackage{algorithmic}
\usepackage{graphicx}
\usepackage{textcomp}
\usepackage{xcolor}
\def\BibTeX{{\rm B\kern-.05em{\sc i\kern-.025em b}\kern-.08em
    T\kern-.1667em\lower.7ex\hbox{E}\kern-.125emX}}

\usepackage{amsmath,amssymb,amsfonts}
\usepackage{booktabs}
\usepackage{xspace}
\usepackage{multirow}
\usepackage{balance}
\usepackage{paralist}
\usepackage{microtype}
\usepackage{listings}
\usepackage{hyperref} 

\usepackage{xcolor}
\usepackage{subcaption}

\RequirePackage{fix-cm}
\usepackage{graphicx}
 \usepackage{subcaption}
\pdfoutput=1
\usepackage{ulem}
\usepackage{enumitem}
\usepackage{wrapfig}
\usepackage{graphics}
\usepackage{graphicx}
\usepackage[graphicx]{realboxes}
\usepackage{rotating}
\usepackage{booktabs}
\usepackage{titlesec}
\usepackage{amsmath,amsfonts,amssymb}
\usepackage{tcolorbox}
\usepackage{multirow}
\usepackage{longtable}
\usepackage{supertabular}
\usepackage{subcaption}
\RequirePackage{float}
\usepackage{epstopdf}

\RequirePackage[skins]{tcolorbox}
\newtcolorbox{custombox}[1]{
	colback=gray!10,
	colframe=gray!70,
	left=1mm,
	right=1mm,
	top=1mm,
	bottom=1mm,
	fonttitle=\bfseries,
	arc=0mm,
	leftrule=1mm,
	rightrule=0mm,
	toprule=0mm,
	bottomrule=0mm,
	notitle,
	before=\par\smallskip\noindent,
	before upper={\textbf{#1: } },
}
\newtcolorbox{probdefinition}[1]{
	colback=gray!10,
	colframe=gray!22,
	left=1mm,
	right=1mm,
	top=1mm,
	bottom=1mm,
	fonttitle=\bfseries,
	arc=2mm,
	leftrule=0mm,
	rightrule=1mm,
	toprule=0mm,
	bottomrule=1mm,
	notitle,
	before=\par\smallskip\noindent,
	before upper={\textbf{#1: } },
}

\usepackage{url} 
\begin{document}

\title{An Empirical Evaluation of White-box and Black-box Test Case Prioritization Techniques in CPSs Modeled in Simulink
}

\author{ Aitor Arrieta \\
  Mondragon University \\
  Mondragon, Spain \\
  \texttt{aarrieta@mondragon.edu}  \\
  }

\date{Received: date / Accepted: date}

\maketitle

\begin{abstract}
MATLAB/Simulink is the leading tool for simulating complex Cyber-Physical Systems (CPSs). The simulation models of complex CPSs are typically compute intensive, and the execution of test cases is long. Furthermore, the execution of test cases is typically triggered several times at different ``in-the-Loop'' test levels (i.e., Model, Software and Hardware-in-the-Loop). Therefore, test optimization techniques, such as test case prioritization, are paramount when testing these systems. In this paper, we present the largest empirical study on test case prioritization techniques for Simulink models by comparing the performance of white-box and black-box test case prioritization techniques. We assess traditional test case prioritization techniques, and we also propose new approaches for use in the context of Simulink models. We empirically compared 11 test case prioritization techniques using six Simulink models of different sizes and complexities. When comparing white-box against black-box test case prioritization techniques, we found that in general, white-box techniques were slightly better than black-box ones. In the context of white-box test case prioritization, the total greedy approach performed better than the additional greedy techniques in larger models. As for the test case prioritization time, black-box techniques were faster, although total greedy techniques were fast enough to be used in practice.

\end{abstract}

\section{Introduction}

Simulation-based testing is a commonly used method for testing and verifying complex and untestable systems such as Cyber-Physical Systems (CPSs)~\cite{Briand2016}. In CPSs and other complex systems where the software interacts with parallel physical processes, simulation models are employed in early stages before implementing the system~\cite{Matinnejad2016}. MATLAB/Simulink is one of the leading tools for modeling and simulating complex dynamic systems in many sectors and domains, including automotive~\cite{Matinnejad2016,Vos2013,Zander-Nowicka2008}, aerospace~\cite{Christhilf2006,Menghi2020}, railway~\cite{Yu2002,Shu2008} and system of elevators~\cite{Sagardui2017a,Sagardui2017}.

While simulation-based testing has several advantages, the execution of test suites in this context can be lengthy in real industrial settings, with single simulations requiring up to an hour to execute~\cite{Menghi2020}. To a large extent, this is caused by the complex mathematical models that are required to model the physical parts of such systems. Other reasons include the need for co-simulation to enable high fidelity simulations that rely on realistic set-ups. For instance, Gladisch et al.~\cite{Gladisch2019} reported that for two autonomous driving case studies (adaptive cruise controller and lane keeping case studies), complex vehicle simulation models were required. Their simulations reduced real-time factors down to 10\%, meaning that 1 minute of simulation time requires 10 minutes to simulate~\cite{Gladisch2019}. In another recent study~\cite{Gonzalez2018}, it was reported that around 8.2 hours were necessary to execute 20 realistic test cases. Besides, such test cases are not executed only once, but several times at different steps and levels, while moving from the simple Model-in-the-Loop (MiL) level to the more complex and realistic simulations that occur at the Hardware-in-the-Loop (HiL) level (see Section \ref{sec:SimBT}). 

To address the high cost of testing CPSs, test optimization methods have relied on several approaches, including test case selection and test case prioritization techniques. Test case selection techniques attempt to select a subset of test cases from a full test suite that are relevant to a specific testing objective; test case prioritization techniques, in contrast, attempt to place test cases in an order that allows them to achieve their objectives (i.e., fault detection) earlier than might otherwise be possible. Arrieta et al.~\cite{Arrieta2018a,Arrieta2019a} studied how different black-box metrics combined with Pareto-based search algorithms performed when selecting a subset of test cases from a test suite. This problem aimed at reducing as much as possible the test execution time while not compromising test quality. However, when using Simulink models it is also important to detect faults as soon as possible. This way, it is possible the debugging process faster~\cite{Menghi2019}. When many test cases exist, many test executions can be performed without finding any faults; thus, effective test case prioritization techniques can be useful. A large corpus of studies have proposed and compared different test case prioritization techniques~\cite{Rothermel1999,Rothermel1997,Rothermel2001,Elbaum2014,Elbaum2002,Epitropakis2015,Hemmati2015,Jiang2015,Jones2003,Korel2005,Korel2008,Luo2016,Luo2019,Shin2019,Li2007} -- although these have been conducted outside of the context of CPSs. Henard et al.~\cite{Henard2016} compared ten white-box and ten black-box test case prioritization techniques, and concluded that there was little difference between the two classes of techniques in terms of rate of fault detection. However, Henard et al.~\cite{Henard2016} focused on unit testing of C programs, which is quite different from testing Simulink models. One major difference is related to semantics. While Simulink models can rely on both discrete and continuous-time semantics, programs written in C often rely only on discrete semantics. This allows the outputs of Simulink models to be connected to their inputs, resulting in continuous feedback over time. On the other hand, coverage-based white-box test case prioritization techniques might not be applicable/effective on some Simulink models due to the difficulty of analyzing coverage for continuous operations~\cite{Matinnejad2019}. Furthermore, in the context of Simulink models, test cases are signals that stimulate the inputs of the system over the time. This provides opportunities to provide novel test metrics based on anti-patterns related to signals, which are among those used in this work. 

In the past, test case prioritization techniques that rely on historical data have performed well, for general purpose software~\cite{Marijan2013a,Wang2016a,Noor2015} and for simulation-based testing models~\cite{Arrieta2019,Arrieta2016a}. However, in order for these techniques to be effective, test cases need to be executed several times~\cite{Khatibsyarbini2017}. To address this problem, test quality metrics that focus either on black-box techniques (i.e., metrics that measure test quality focusing solely on the inputs and outputs of the simulation models) or on white-box techniques (i.e., metrics that measure which parts of the simulation models have been exercised) can be employed. To the best of our knowledge, this is the first paper in which different white-box and black-box test case prioritization techniques are proposed and empirically evaluated in the context of MATLAB/Simulink models. 

Specifically, this paper makes the following key contributions. First, we present and revisit test case prioritization techniques for the context of MATLAB/Simulink models. Second, we conduct an empirical evaluation considering six subject models of different sizes and complexities and 11 test case prioritization techniques. We also compared these techniques with a baseline test case prioritization approach and with an optimal test case ordering. To the best of our knowledge, this is the largest test prioritization empirical evaluation performed to date in the context of Simulink models. Third, we provide implementations of all of the techniques that we evaluated in a public repository that can be accessed by other researchers and practitioners. We further provide all of the evaluation materials, including scripts, models, mutants, test cases and sources for statistical analysis in a public repository to support replicability and reproducibility.\footnote{Replication package can be found in \url{https://doi.org/10.6084/m9.figshare.28777829}}

The main findings of our study can be summarized as follows. First, we found that for Simulink models, unlike in the context of multi-objective test case selection~\cite{Arrieta2019a}, white-box metrics are generally as competitive as black-box metrics in the context of test case prioritization for Simulink models. Second, we found that there is no single technique that stands out over the rest of the techniques considered, although in most cases, all techniques outperformed the selected baseline technique. Third, in the context of white-box test case prioritization, the total greedy based techniques perform better than the additional greedy-based techniques used in 50\% of the used subject models. Total and greedy-based test case prioritization algorithms are a class of techniques that utilize feedback about coverage data achieved by test cases prioritized thus far under-performs a class of techniques that do not. Fourth, executing Mathworks' API for measuring white-box coverage in MATLAB for Simulink models is time consuming, and significantly increases the running time of white-box test case prioritization techniques.

\section{Background}
\label{sec:background}
\subsection{Simulation models and basic notation}
\label{sec:SimulationModels}

\begin{figure}[h]
	\centering
	\includegraphics[trim = 0 30 320 0,clip, width=0.65\linewidth]{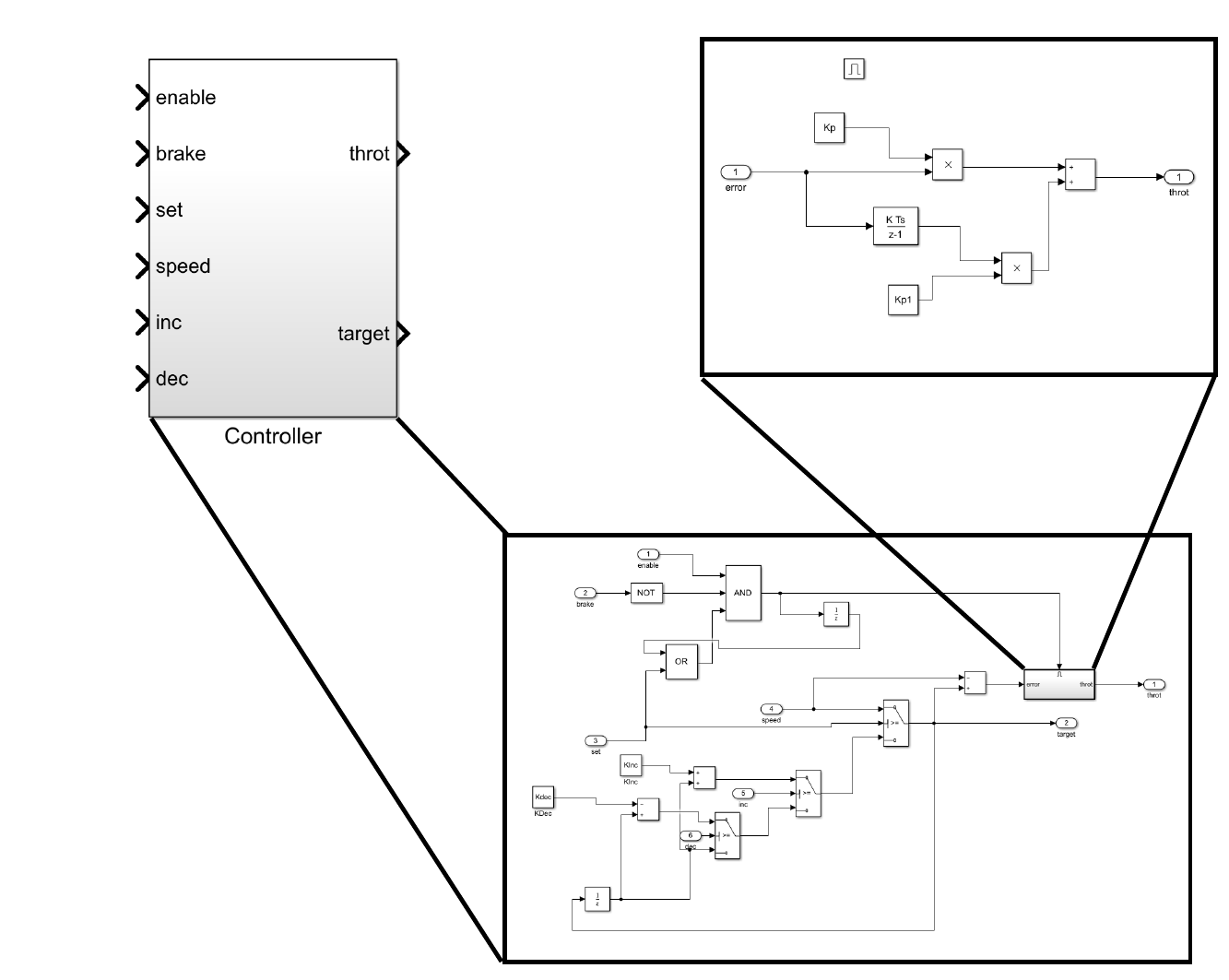} 
	\caption{Example of a simulation model of a Cruise Controller of a car~\cite{Arrieta2019a}}
	\label{fig:SimulationModel}
	\hfill	
\end{figure}

Simulation is the process of creating a digital model of a system in order to be able to predict the performance of an actual system in the real world~\cite{Arrieta2019a}. CPS developers rely on simulation tools and Model-Based Design (MBD) work-flows, where graphical models of their systems are employed to do rapid prototyping~\cite{Chowdhury2018}. Simulation tools oriented to the design and development of CPSs (e.g., Simulink) are often dataflow models, where each model contains a set of blocks~\cite{Chowdhury2018}. Each block from these models accepts data through its inputs and may pass output through its output ports after a set of operations (e.g., mathematical, logical) are performed. Figure \ref{fig:SimulationModel} provides an example of a simulation model, which includes six inputs (enable, brake, set, speed, inc and dec) and two outputs (throt and target). 

\label{sec:BasicNotation}

\begin{figure}
    \centering
    \includegraphics[trim= 120 330 120 330,clip, width=\linewidth]{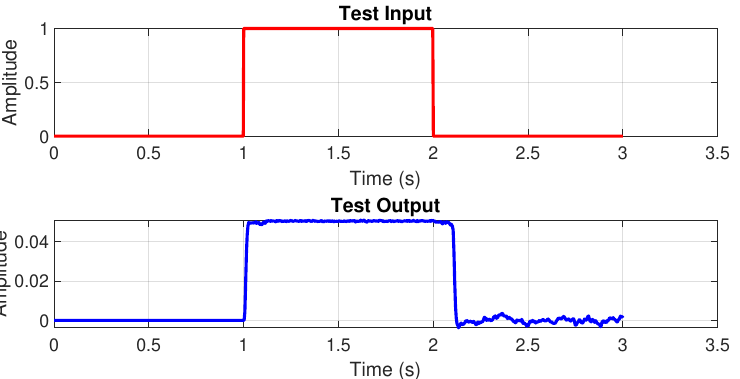}
    \caption{An example of the execution of a test case in the EMB case study system}
    \label{fig:example}
\end{figure}

The remainder of this paper makes use of the following definitions and notation. We note that, similar to previous studies~\cite{Matinnejad2016,Matinnejad2019,arrieta2023some,Arrieta2019}, we assume that the employed simulation employs a fixed sample step. Let $SM=(I,O)$ be a simulation model, where $I=\{i_1,i_2, ..., i_N\}$ is a subset of inputs and $O = \{o_1, o_2, ..., o_M\}$ is a subset of outputs~\cite{Matinnejad2016}. Each input and output of the simulation model is a signal (i.e., a function of time), which is stored as a vector where elements are indexed by time~\cite{Matinnejad2016}. Figure \ref{fig:example} shows the input and corresponding output of a test case for one of the case study systems used in our evaluation. The simulation time ($T$) is divided into a set of equal sample time steps ($\Delta T$)~\cite{Matinnejad2016}. A signal ($sig$) is a function in time of a set of $k$ observed simulation steps (i.e., $ sig: \{0, \Delta T, 2 \times\Delta T , ..., k\times \Delta T\} $). For instance, a simulation of 10 seconds (i.e., T=10), with a sample time of 0.05 seconds (i.e., $\Delta T$ = 0.05) would have a total of 10/0.05+1 (i.e., k=201) simulation steps. The lower the sample time, the higher the precision of the simulation. However, the time required to simulate the system will also be lengthier. In our study, for each model being tested, we consider a fixed simulation sample time (i.e., the same for all the test cases).

\subsection{Simulation-based testing}
\label{sec:SimBT}

The testing of CPSs occurs across several levels.
These levels, defined as Model-in-the-Loop (MiL), Software-in-the-Loop (SiL), Processor-in-the-Loop (PiL) and Hardware-in-the-Loop (HiL), are typically employed in the CPS development work-flow at different verification and validation stages. In the first stage, when engineers are building the model of the CPS, the MiL test level is employed. MiL is the basic simulation that engineers employ to analyse the embedded software's model with the physical plant of the CPS~\cite{Shokry2009}. At this level, the software is a model and computations are performed using floating-point arithmetic to obtain simulation results as reference values~\cite{Shokry2009}. These reference values are later used and compared in the following simulations. When the test results are considered satisfactory at the MiL test level, the code of the controller is generated and the software's model is replaced with executable code (e.g., a *.dll)~\cite{Shokry2009}. At this test level, fixed-point computations are employed, as they will in the real target processor. At the following test level (i.e., PiL), the code is cross-compiled and executed on the real target processor, which communicates with the simulation tool that simulates the physical layer of the CPS in the computer~\cite{Shokry2009}. The main objective of this test level is to detect potential faults introduced by the compiler~\cite{Shokry2009}. In many occasions, the PiL test level is avoided and the HiL test level is next employed~\cite{ayerdi2020towards}. At the HiL test level, the software is integrated with the Electronic Control Unit (ECU) and all the real-time infrastructure (including drivers, real-time operating systems, communications, etc.)~\cite{Shokry2009}. The physical layer is encapsulated and emulated in an embedded device (e.g., FPGA) and the simulation is performed in real-time~\cite{Shokry2009}. At this level, in addition to testing functional requirements, non-functional requirements are also validated, because temporal constraints are critical for CPSs. Several different approaches have been proposed for testing CPSs at different test levels, from MiL and SiL~\cite{Matinnejad2015a,Matinnejad2016,Matinnejad2019,Arrieta2017,Arrieta2016,Sagardui2017a,Sagardui2017} to HiL~\cite{Eidson2011,Kane2014}.

In our experience with industrial partners, we have discovered that test cases are not executed only once at each level, but several times. Furthermore, the software testing phases for CPSs are carried out in a scaled manner. The initial model is usually low fidelity and simple, but as functionalities are validated, simple subsystems in the CPS model are replaced by more complex ones. At the HiL test level a similar process occurs. The ECU includes other functionalities and tasks apart from the functionalities that are tested at the MiL and SiL test levels. At this stage, test cases are executed several times, because there are many changes required (e.g., changes in task priorities, addition of functionalities not tested in previous stages, changes in communication protocols, changes in the configuration of the operating system, and so forth). Thus, at all test levels, regression test optimization approaches play an important role.

\section{Selected Test Case Prioritization Methods}
\label{sec:TPM}

In this section we explain the adapted test case prioritization techniques and how we used them in the context of Simulink models for CPSs testing. We first explain the adapted black-box techniques (Section \ref{sec:bbt}) followed by the white-box techniques (Section \ref{sec:wbt}).%

\subsection{Black-Box Techniques}
\label{sec:bbt}

Black-box techniques are of special interest in the context of simulation-based testing, as white-box metrics are not always available. This might be due to several reasons. For instance, developers of CPSs typically involve several Original Equipment Manufacturers (OEMs), from different engineering domains (e.g., mechanical, electrical, computer science, etc.). Each of them provide their system models as a functional black-box model without access permission by other OEMs due to Intellectual Property (IP) issues. Subsequently, it is often infeasible to measure white-box coverage. Another reason is that complex software of Simulink models is typically developed in C, and later executable code is generated and incorporated as a subsection in Simulink. This executable code is also black-box and it is not possible to obtain coverage information. The proposed black-box techniques for test case prioritization can be catalogued into two main categories: (1) techniques based on anti-patterns and (2) techniques based on test similarity. These techniques were first proposed by Matinnejad et al.~\cite{Matinnejad2016} for test case generation and later tailored by Arrieta et al.~\cite{Arrieta2019a} for regression test optimization. We now explain in detail each of them.

\subsubsection{Techniques based on anti-patterns}

By stimulating the inputs of a simulation model, it is possible to obtain the output results of the simulation and provide some numerical values with respect to certain anti-patterns. In previous works~\cite{Matinnejad2019,Matinnejad2016,Matinnejad2017} three different anti-patterns were defined for Simulink models, which we later adapted to the context of test case selection~\cite{Arrieta2018a,Arrieta2019a}. The anti-pattern metrics quantitatively measure certain patterns that are often related with incorrect/faulty behaviours of simulation models~\cite{Matinnejad2019,Matinnejad2016,Matinnejad2017}. 
The hypothesis behind these measures is that the higher the anti-pattern degree a test case has, the higher the likelihood of detecting a fault. We thus sort the test cases in a descending order based on its degree of anti-pattern, which are explained in detailed below.

\subsubsection*{Metrics for use in prioritization techniques based on anti-patterns:}

For the context of Simulink models, three anti-patterns were identified in previous works~\cite{Matinnejad2016,Matinnejad2017,Matinnejad2019}, which are measured by considering the outputs signals of the models. These anti-patterns are named (1) instability, (2) discontinuity and (3) growth to infinity. These anti-patterns were used as black-box metrics to prioritize test cases in the context of Simulink models. 

\textbf{Instability: }Instability measures the degree at which an output signals shows quick and frequent oscillations~\cite{Matinnejad2017}. An example of when this anti-pattern appears is when a state-chart controller quickly switches between states. This effect is (generally) undesirable in (most) physical processes of CPSs~\cite{Matinnejad2017}. For instance, having this kind of patterns in the acceleration or the speed of an autonomous vehicle is not desirable. The instability of a signal output $sig$ can be measured by applying Equation \ref{eq:instability}, where $k$ is the number of simulation steps and $\Delta T$ is the simulation time step. The higher the value of $instability(sig)$, the higher the instability degree of the signal.\footnote{We consider the notation from Section~\ref{sec:SimulationModels} to explain the equations of this section}

\begin{equation}
\label{eq:instability}
instability(sig) =  \sum_{i=1}^{k}|sig(i\cdot\Delta T)-sig((i-1)\cdot\Delta T)|
\end{equation}

Simulink models usually have more than one output, and therefore, the instability degree of a test case needs to be normalized. In our approach, we assume that there is no domain knowledge, and thus, we give the same importance to all output signals. Given an $SM=\{I,O\}$ with $M$ outputs (i.e., $O = \{o_1, o_2, ..., o_M\}$), the instability of a test case $j$ in $TS$ is measured with Equation \ref{eq:instabilityTC}, where $instability(Osig_{j_i})$ is the instability of the $i$-th signal for $tc_j$ and $\max(instability(Osig_i))$ is the maximum instability degree obtained in the $i$-th signal when considering all test cases in $TS$.

\begin{equation}
\label{eq:instabilityTC}
TCInstability(tc_j)=\frac{\sum_{i=1}^{M}instability(Osig_{j_i})}{\sum_{i=1}^{M}\max(instability(Osig_i))}
\end{equation}

\textbf{Discontinuity: }Discontinuity refers to an anti-pattern where an output signal displays a short duration pulse~\cite{Matinnejad2017}. This anti-pattern might appear, for instance, when an unforeseen situation from the environment arises and the system needs to self-regulate to adapt to the situation. This anti-pattern can be measured at a signal level by computing its derivative function~\cite{Arrieta2019a}, but as simulation time steps are not infinitesimal, Matinnejad et al. proposed a function that relies on discrete change rates~\cite{Matinnejad2015b}. Given a signal $sig$, $lc_i = |sig(i\cdot \Delta t) - sig((i-dt)\cdot \Delta t|/\Delta t$ is the left change rate at step $i$, and $lc_i = |sig((i+dt)\cdot \Delta t-sig(i\cdot \Delta t)|/\Delta t$ is the right change rate at step $i$~\cite{Matinnejad2015b,Matinnejad2017}. The discontinuity of a signal $sig$ can be measured by applying Equation \ref{eq:discontinuity}, i.e., the maximum of the minimum of the left and right change rates at each simulation step for a given set of simulation steps~\cite{Matinnejad2015b,Matinnejad2017}.

\begin{equation}
\label{eq:discontinuity}
discontinuity(sig)=\max\limits_{dt=1}^{3}(\max\limits_{i=dt}^{k-dt}(\min(lc_i,rc_i)))
\end{equation}

To normalize the discontinuity degree of each test case for more than one output, the discontinuity of a test case $j$ in $TS$ is obtained with Equation \ref{eq:discontinuityTC}. For a test case $j$, given an $SM=\{I,O\}$ with $M$ outputs (i.e., $O = \{o_1, o_2, ..., o_M\}$), the discontinuity degree $discontinuity(Osig_{j_i})$ is the discontinuity of the $i$-th signal for $tc_j$ and $max(discontinuity(Osig_i))$ is the maximum discontinuity value in the $i$-th signal when considering all test cases in $TS$.

\begin{equation}
\label{eq:discontinuityTC}
TCDiscontinuity(tc_j) = \dfrac{\sum_{i=1}^{M}discontinuity(Osig_{j_i})}{\sum_{i=1}^{M}max(discontinuity(Osig_i))}
\end{equation}

\textbf{Growth to infinity: }Another anti-pattern for simulation models 
can be the growth to infinity, which measures the maximum absolute value of a signal. This metric can be effective as, for instance, the control algorithms of an hybrid system (e.g., a CPS) can be wrongly tuned. When this happens, the feedback control system can be saturated, leading the system to an uncontrolled state and resulting in signals of huge values. The anti-pattern ``Growth to infinity'' of a signal $sig$ can be measured by obtaining its maximum absolute value across all sample steps (Equation \ref{eq:infinite}).

\begin{equation}
\label{eq:infinite}
inf(sig) =  max|sig(i\cdot\Delta T)|
\end{equation}

As for the previous anti-patterns, a normalization process is required to account for all the output signals of a simulation model. Given an $SM=\{I,O\}$ with $M$ outputs (i.e., $O = \{o_1, o_2, ..., o_M\}$), the growth to infinity of a test case $j$ in $TS$ is obtained with Equation \ref{eq:infiniteTC}, being $inf(Osig_{j_i})$  the growth to infinity of the $i$-th signal for $tc_j$ and $max(inf(Osig_i))$ the maximum infinity value in the $i$-th signal when considering all test cases in $TS$.

\begin{equation}
\label{eq:infiniteTC}
TCInfinity(tc_j) = \dfrac{\sum_{i=1}^{M}inf(Osig_{j_i})}{\sum_{i=1}^{M}max(inf(Osig_i))} 
\end{equation}

\subsubsection{Techniques based on test similarity metrics}
\label{sec:similarityTechniques}

The hypothesis that dissimilar test cases have larger probabilities of finding faults has been widely investigated~\cite{Feldt2016,Chen2004,Chen2004a,Chen2010,Hemmati2010,Hemmati2013}. In the context of test case prioritization, information on how similar two test cases are can be used to prioritize test cases. Similarity between two test cases can be measured by a distance function, where a small distance means that two test cases are similar whereas a large distance means the opposite. With a distance metric, it is possible to build a distance matrix to measure similarity among all the test cases in the test suite. This information is later used by the test case prioritization algorithm to iteratively include a test case in the prioritized test suite. In each iteration, the algorithm includes the test case which is farthest from those test cases already included in the prioritized test suite. Several distance metrics have been proposed in the literature to measure similarity between test cases (e.g., Hamming, Yaccard)~\cite{Hemmati2010}. Based on previous studies~\cite{Matinnejad2015b,Matinnejad2016}, we selected the Euclidean distance to measure the similarity between two test cases. Since these metrics can be applied both on inputs as well as on outputs of the simulation models, we applied it on both, yielding two similarity-based test prioritization techniques, as explained below.

\subsubsection*{Metrics for test prioritization techniques based on similarity:}

 Two metrics were derived to prioritize test cases based on test similarity. 
 The former measures the similarity of test cases by considering the stimulation signals, whereas the latter measures it by considering the outputs. As previously mentioned, the Euclidean distance is employed to measure the similarity of two test cases, as proposed in previous studies~\cite{Matinnejad2015b,Matinnejad2016,Arrieta2018a,Arrieta2019a}. In the context of Simulink models, test cases can have different test execution times, and subsequently, a different number of simulation steps. In prior studies the Euclidean distance was adapted to consider this~\cite{Arrieta2018a,Arrieta2019a}.

 The Euclidean distance to measure the distance between two signals related to a specific input or output (i.e., $sig$ and $sig'$) in two different test cases can be measured by applying Equation \ref{eq:EuclideanDistance}. For both signals, $\min(k_{sig},k_{sig'})$ is the number of steps of the signal whose test case has a lower Test Execution Time (TET), $max_{R_{sig}}$ is the maximum value that the signal in the simulation can obtain, $min_{R_{sig}}$ is the minimum value that the signal in the simulation can obtain and $K$ is the number of steps that the test case with the highest TET in the test suite has~\cite{Arrieta2018a,Arrieta2019a}. As in prior work~\cite{Arrieta2018a,Arrieta2019a}, $K$ is considered as the number of steps for the longest test case for two main reasons. Firstly, it ensures that the distances between test cases are normalized. Secondly, because a longer test case might have a higher chance to detect faults. Therefore, the distance of those very short test cases is penalized. It is important to note that the simulation step ($\Delta t$) is considered the same for all test cases. A higher Euclidean distance means that two signals are less similar.

 \begin{equation}
 \label{eq:EuclideanDistance}
 D(sig,sig') = \\
 \dfrac{\sqrt{\sum_{i=0}^{\min(k_{sig},k_{sig'})}(sig(i \cdot \Delta t)-sig'(i \cdot \Delta t))^{2}}}{\sqrt{K+1}\times (max_{R_{sig}}-min_{R_{sig}})}
 \end{equation}

\textbf{Input-based test similarity: }Given two test cases ($TC_a$ and $TC_b$), for a simulation model with N inputs, Equation \ref{eq:distanceTC} defines the input signal-based distance, which is the sum of the normalized Euclidean distance of each input signal between both test cases. A higher distance means that the test cases are less similar in terms of their input signals.

\begin{equation}
\label{eq:distanceTC}
InTCD(TC_a,TC_b) = \sum_{i=1}^{N}D(Isig_{a_{i}},Isig_{b_{i}})
\end{equation}

\textbf{Output-based test similarity: }Given two test cases ($TC_a$ and $TC_b$), for a simulation model with N outputs, Equation \ref{eq:distanceTC_out} defines the output signal-based distance, which is the sum of the normalized Euclidean distance of each output signal between both test cases. A higher distance means that the test cases are less similar in terms of their output signals.

\begin{equation}
\label{eq:distanceTC_out}
OutTCD(TC_a,TC_b) = \sum_{i=1}^{N}D(Osig_{a_{i}},Osig_{b_{i}})
\end{equation}

\subsection{White-Box Techniques}
\label{sec:wbt}

Two variants of traditional ``greedy'' dynamic test case prioritization techniques are the most common test prioritization strategies when using white-box metrics: the total strategy and the additional strategy.\footnote{A greedy algorithm is a technique that makes next decisions based on local optima without considering subsequent backtracking}  On the one hand, the total greedy technique prioritizes test cases based on their absolute code coverage~\cite{Luo2019}. Specifically, supposing that statement coverage is used as a proxy to measure the test quality, tests are sorted in descending order based on the number of test objectives covered by each test case. On the other hand, the additional greedy strategy focuses on test objectives (e.g., statements) not yet covered by test cases already prioritized. At each iteration, the additional greedy algorithm selects the test case that covers more statements that have not been covered yet. When all the test objectives have been covered or no additional test objective can be covered again, these objectives are reset and the test prioritization process begins again\textcolor{black}{~\cite{li2021aga}}. 

Both of them have different strengths and weaknesses~\cite{Hao2014}. As explained by Hao et al.~\cite{Hao2014}, the negative aspect of the total technique could be that if a test case covers a set of statements that only includes a fault, and that test case detects that fault, it is unnecessary to consider the same statements again for the remaining test cases. However, the covering of a statement does not ensure the detection of all faults. If the fault is not detected by the test case, but it is detectable by others, it is important to consider those statements. In summary, when using the total strategy, the detection of faults in statements that are covered infrequently may be delayed~\cite{Hao2014}. On the other hand, for the additional technique, as the strategy considers not covered objectives, it avoids the problem exhibited by the total greedy technique, where statements that already exhibited all the possible faults are considered again. However, if a fault is not detected in a statement covered by a test case, the detection of this may be delayed. In summary, the risk of using the additional technique is that the detection of faults may be delayed in statements that are covered frequently~\cite{Hao2014}.


Different coverage metrics have been investigated to use as a surrogate of test objectives and at different granularity levels (i.e., line, statement, method class, etc.). For this study, both techniques were used, i.e., total and additional. Each of these techniques was integrated with three white-box coverage metrics: Decision Coverage (DC), Condition Coverage (CC) and Modified Condition/Decision Coverage (MC/DC). These three metrics have been used because they are available in the model coverage API provided by the Mathworks.\footnote{https://mathworks.com/help/slcoverage/}

 \subsection{Test case prioritization techniques not selected in this study}
 \label{sec:notSelected}

 In the literature, different techniques and metrics have been defined but do not fit into our study. In this section we explain why these techniques were discarded in our study. On the one hand, generally, the so-called static methods~\cite{Luo2019} are not applicable in the context of Simulink models. For instance, the call-graph-based static test case prioritization approaches make use of test code to extract which methods are invoked by each of them~\cite{Luo2019,Zhang2009b}, either in a total way or in an accumulative way. Other static approaches include string-distance based, where similarities of test cases are measured based on string edit distances~\cite{Ledru2012}. There are other static approaches (e.g., topic-based), which make use of test code information

 On the other hand, there are some dynamic test case prioritization approaches that could not be selected for our evaluation. Some regression test optimization approaches that rely on requirements covered by test cases~\cite{Lachmann2017} have also been applied to test case prioritization of CPSs modeled in Simulink~\cite{Arrieta2019,Arrieta2016b}. Another effective way for regression test case prioritization is the use of historical data (e.g., number of faults detected by test cases) in order to effectively prioritize test cases~\cite{Arrieta2016a,Arrieta2019,Wang2014}. However, all this information is typically not available, specifically for those cases where Simulink Design Verifier or other Simulink test generation tools (e.g.,~\cite{Matinnejad2016,Arrieta2018}) are used to generate test cases. The idea of the selected techniques in this paper is that the test case prioritization techniques can be easily applied in any Simulink model. In the case of history-based test case prioritization, these have been found to require large start-up information in order to be effective~\cite{Khatibsyarbini2017}.
 
Adaptive Random Test Case Prioritization (ART) is a dynamic technique that it randomly selects a set of test cases iteratively to build a candidate set, and from this candidate set it selects the farthest test case (based on the distances) from the already prioritized test cases. It is similar to the selected test case prioritization techniques based on the similarity, but the main difference is that instead of picking the farthest test case from all the non-prioritized test cases, it picks the farthest test case from a subset of the non-prioritized test cases which are randomly selected. An advantage with respect to the techniques proposed in Section \ref{sec:similarityTechniques} is that it might be faster as it does not require to measure the distance between all the non-prioritized test cases and all the prioritized ones. Nevertheless, after implementing our distance-based test case prioritization techniques, we figured out that it was fast enough to be used in practice (i.e., less than 0.02 seconds), and thus, this technique is more reliable than the ART, as it measures the distance of all the non-prioritized test cases. 
 
Many approaches make use of mutation scores as a metric to assess the quality of test cases~\cite{Henard2016,Henard2013,Papadakis2014,Lou2015,Shin2019}. However, mutation testing is a very expensive technique in the context of CPSs~\cite{cornejo2021mutation,vigano2022data}, including Simulink models because, besides the control algorithms, they typically involve the physical layer of the model~\cite{Arrieta2019,Arrieta2018a,Arrieta2019a}. This physical layer is usually developed with complex mathematical models \textcolor{black}{, such as differential equations. Solving complex differential equations that represent the dynamics of physical processes require significant computational resources and time, as the differential equations must be numerically integrated over time to produce accurate results. Each mutant introduces a variation that requires a full re-simulation to observe its effects, significantly increasing the computational load and making this technique not scalable in this context. Thus, the use of mutants to assess the adequacy of individual test cases is unfeasible in practice. }

\section{Empirical Evaluation}
\label{sec:evaluation}

In this section we explain the carried out empirical evaluation of the proposed test case prioritization strategies in the previous section. Table \ref{tab:TechniquesTables} summarizes the employed black-box and white-box techniques and provides an acronym of each strategy.

\begin{table}
	\centering
    \scriptsize
	\caption{Summary table for evaluated black-box and white-box test prioritization techniques}
	\label{tab:TechniquesTables}
	\begin{tabular}{llll}
		\hline
		\multicolumn{1}{c}{\textbf{Acronym}} & \multicolumn{1}{c}{\textbf{Strategy}} & \multicolumn{1}{c}{\textbf{Metric}} & \multicolumn{1}{c}{\textbf{Type}} \\ \hline
		AP-Ins                                 & Anti-patterns based                    & Instability                          & \multirow{5}{*}{Black-Box}         \\ 
		AP-Disc                                & Anti-patterns based                    & Discontinuity                        &                                    \\ 
		AP-GTI                                 & Anti-patterns based                    & Groth to infinity                    &                                    \\ 
		SB-IS                                  & Similarity-based                       & Input similarity                     &                                    \\ 
		SB-OS                                  & Similarity-based                       & Output similarity                    &                                    \\ \hline
		Add-DC                                  & Additional                & Decision Coverage                    & \multirow{6}{*}{White-Box}         \\ 
		Add-CC                                  & Additional                & Condition Coverage                   &                                    \\ 
		Add-MCDC                                & Additional                 & Modified Condicion/Decision coverage &                                    \\ 
		Tot-DC                                 & Total                                  & Decision Coverage                    &                                    \\ 
		Tot-CC                                 & Total                                  & Condition Coverage                   &                                    \\ 
		Tot-MCDC                               & Total                                  & Modified Condicion/Decision coverage &                                    \\ \hline
		Sanity Check                               & Similarity-based                       & Input similarity (minimization)          & Black-Box                          \\ \hline
	\end{tabular}
 \vspace{-0.55cm}
\end{table}

\subsection{Research Questions}

To compare the selected test case prioritization methods explained in Section \ref{sec:TPM} in the context of Simulink models, we addressed the following Research Questions (RQs):

\textit{\textbf{RQ1 -- Sanity check:} How well do the selected techniques perform with respect to the sanity check (SC) technique when considering the fault detection rate?} This RQ was defined as a sanity check to verify that, on the one hand, the test case prioritization problem in the context of Simulink models is a non-trivial problem, and, on the other hand, to ensure that the proposed techniques are adequate by comparing them against a baseline. For the sanity check, as a baseline, similar to~\cite{Henard2016}, we used the prioritization based on diversity of test cases. However, instead of maximizing the diversity, the algorithm minimized the diversity of test cases based on the Euclidean distance. 

\textit{\textbf{RQ2 -- Black-box techniques:} How well do the studied black-box prioritization techniques perform in terms of fault detection rate?} We defined this RQ with the aim of comparing the performance of the different black-box test case prioritization techniques. The aim was to determine whether there is a technique that stands out over the rest in order to be proposed to practitioners for the cases where there is no access to white-box data. RQ2 is especially important in the context of CPSs, as not having access to the code is a common scenario. Additionally, it is a common scenario in Simulink models that there are not many routing/flow control elements, such as switches or state-flow charts. Information of white-box coverage may not be sensitive enough when not having these types of elements available. 
	
\textit{\textbf{RQ3 -- White-box techniques:} How well do the studied white-box prioritization techniques perform in terms of fault detection rate?} White-box test quality metrics have been traditionally employed for software testing. However, few studies have assessed their performance in the context of Simulink models. In some cases, it is possible to measure the white-box coverage of the models, and thus, it can be used as the metric for test case prioritization. Similar to RQ2, RQ3 aimed at comparing the performance of traditional white-box test case prioritization techniques. The goal is to determine whether there is a clear winner among the selected white-box test case prioritization techniques or whether this depends on the subject model. 

\textit{\textbf{RQ4 -- Best technique:} How well do the black-box techniques compare with the white-box ones in terms of fault detection rate? How far are the techniques compared with the optimal test case prioritization technique in terms of fault detection rate? }From RQ2 and RQ3, for each of the used subject models, the best white-box and black-box techniques were obtained. The fourth RQ aimed at comparing the best black-box technique with the best white-box one for each subject model with the aim of determining the best overall approach. We also considered an optimal ordering of the test cases for each case study to provide an upper bound for the comparison~\cite{Elbaum2002}. This was obtained by considering the information of test cases exposing the faults. This is not feasible in practice, as the relation between faults and test cases is unknown, but it helps us gain insight into the success rate of the selected techniques~\cite{Rothermel1999}. To do this, we employ an additional greedy technique considering mutants as test objectives (i.e., instead of test coverage). The additional greedy strategy focuses on mutants not yet killed by test cases that have already been included in the prioritization queue. At each iteration, the additional greedy algorithm selects the test case that covers more mutants not killed yet by prioritized test cases.  Notice that this comparison has been performed in other similar empirical studies (e.g.,~\cite{Elbaum2002,Rothermel1999,Zhang2013b}).

\textit{\textbf{RQ5 -- Test case prioritization time:} How do black-box and white-box techniques compare in terms of the required execution time for the prioritization process?} In situations where very little time is available for the overall regression testing process (e.g.,~\cite{Gligoric2014}), the time it takes each technique to return a prioritized test suite becomes an important factor~\cite{Henard2016,Gligoric2014}. The fifth RQ aims at studying which technique would be the most appropriate in situations where there is a limited test execution time budget.

\subsection{Selected Benchmark}

Besides investigating the performance of black-box and white-box test prioritization techniques in the context of Simulink models, we also wanted to contrast the findings of our work with those from the context of test case selection~\cite{Arrieta2019a,Arrieta2019}. To make our comparison as fair as possible, we used as much as possible the benchmark prepared by Arrieta et al.~\cite{Arrieta2019a,Arrieta2019}.

\subsubsection{Subject Models}

Six Simulink models of different sizes and complexities were employed to compare the selected techniques. The key characteristics of the selected subject models are summarized in Table~\ref{tab:CaseStudy}, which includes the number of inputs, outputs and blocks. All these subjects have been previously used in similar studies for evaluating testing methods~\cite{Arrieta2019a,Arrieta2019,Menghi2019,Matinnejad2016,Matinnejad2017,Matinnejad2019,Nejati2019}.

Five out of the six subject models were the same as those employed by Arrieta et al.~\cite{Arrieta2019a,Arrieta2019} in their benchmark. The Car Window (CW) subject involves an open-source system that models four car windows. The physical layer involves the electrical and mechanical models of the four windows. Each window is controlled by a subsystem that also involves a state machine. The Electro-Mechanical Braking (EMB) system is an industrial open source model developed by Bosch~\cite{Strathmann2015}. This system combines physical and software models, with the software model controller including a discrete state machine and a continuous PID controller~\cite{Matinnejad2017}, and it has been used to evaluate the performance of a test generation algorithm~\cite{Matinnejad2017}. The Cruise Controller (CC) model involves only the controller of a system and it does not contain any plant subsystem. This model was used in~\cite{Matinnejad2016} to assess a test case generation approach for Simulink models. The AC Engine model involves an AC engine in combination with a controller that also includes some safety functionalities~\cite{Arrieta2019,Arrieta2018,Arrieta2017}. The Two Tanks model involves a system where a controller regulates the incoming and outgoing flows of two tanks. This subject was used to evaluate a test oracle generation approach~\cite{Menghi2019}, a comparison between model testing and model checking approaches~\cite{Nejati2019} or test case selection approach based on multi-objective search algorithms~\cite{Arrieta2019}. In the original benchmark, Arrieta et al.~\cite{Arrieta2019a,Arrieta2019} employed an additional subject model involving a toy artificial simulink model named \textit{``Tiny''}, which only involved 15 blocks. Because the size of this model is very small to draw any meaningful conclusions, similar to other studies (e.g.,~\cite{ling2023not}), we decided to remove that subject model from the benchmark, and instead include an additional one, i.e., Swimairspeed. Swimairspeed is one model from ``the ten lockheed martin cyber-physical challenges'' that involves a safety algorithm for monitoring airspeed in the system wide integrity monitor suite~\cite{mavridou2020ten}; the algorithm provides a warning to an operator when the vehicle speed is approaching a boundary where an evasive flyup maneuver cannot be achieved~\cite{mavridou2020ten}.

\subsubsection{Test Cases}

For the five models we reused from the initial benchmark proposed by Arrieta et al.~\cite{Arrieta2019a,Arrieta2019}, we employed the exact same test cases. For the EMB, CC and Two Tanks models, Arrieta et al.~\cite{Arrieta2019a,Arrieta2019} generated 150 test cases randomly. This was due to these models not being compatible with Simulink Design Verifier. For the CW case study,  Arrieta et al.~\cite{Arrieta2019a,Arrieta2019} used Simulink Design Verifier (SDV) to automatically generate test cases following an MC/DC white-box coverage criterion, yielding a total of 133 test cases given 100,000 seconds of test generation budget. For the AC Engine case study, Arrieta et al.~\cite{Arrieta2019a,Arrieta2019} reused 120 test cases that were generated with a test case generation tool from a previous study~\cite{Arrieta2017}. For the Swimairspeed model, we had to generate test cases from scratch. After adapting the model configuration, we employed SDV. 3 of the test cases were generated by establishing the decision-condition coverage criterion. 52 test cases were generated following the MC/DC coverage criterion. To have a total of 150 test cases, we complemented these test cases by generating 95 additional test cases randomly. The size of the test suites is aligned with typical industrial CPSs (e.g., 50 to 150 test cases for a CPS for a satellite system~\cite{Shin2018}). Table \ref{tab:CaseStudy} reports the key characteristics of the employed test cases too, including the obtained coverage as well as the execution time. It is important to highlight two key considerations. First, this execution was conducted at the Model-in-the-Loop (MiL) testing level. Even at this stage, variations in execution may arise due to differences in the fidelity of the physical plant representation within the Cyber-Physical System (CPS). Second, these executions must be extended to higher testing levels, namely Software-in-the-Loop (SiL) and Hardware-in-the-Loop (HiL). In particular, based on our experience with industrial partners, many tests at the HiL stage necessitate manual execution. Furthermore, testing at this level may involve risks of hardware damage~\cite{Shin2018}. Consequently, test prioritization is crucial not only from the test execution time context.

\begin{table}[t!]

	\centering
    \tiny
	\caption{Key characteristics of the selected case studies}
	\label{tab:CaseStudy}

\begin{tabular}{lrrrrlrrrrr}
\hline
\textbf{Case Study} & \multicolumn{1}{l}{\textbf{Blocks}} & \multicolumn{1}{l}{\textbf{Inputs}} & \multicolumn{1}{l}{\textbf{Outputs}} & \multicolumn{1}{l}{\textbf{TCs}} & \textbf{Test Generation} & \multicolumn{1}{l}{\textbf{Mutants}} & \multicolumn{1}{l}{\textbf{DC Cov}} & \multicolumn{1}{l}{\textbf{CC Cov}} & \multicolumn{1}{l}{\textbf{MCDC Cov}} \\ \hline
CW                  & 235                                  & 15                                   & 4                                     & 133                                      & SDV (MC/DC)              & 96                                    & \textcolor{black}{86.41\%}                              & \textcolor{black}{86.66\% }                             & \textcolor{black}{73.21\% }                                                             \\ 
EMB                 & 315                                  & 1                                    & 1                                     & 150                                      & Random                   & 18                                    & \textcolor{black}{82.69\%  }                            & \textcolor{black}{100\%}                                & \textcolor{black}{66.67\% }                                                               \\ 
CC                  & 31                                   & 5                                    & 2                                     & 150                                      & Random                   & 20                                    & \textcolor{black}{100\% }                               & \textcolor{black}{100\%  }                              & \textcolor{black}{100\%}                                                                 \\ 
Swimairspeed        & 130                                  & 7                                    & 5                                     & 150                                      & \textcolor{black}{SDV + Random}             & 9                                     & \textcolor{black}{100\%  }                              & \textcolor{black}{100\% }                               & \textcolor{black}{100\% }                                                           \\ 
ACEngine            & 257                                  & 4                                    & 1                                     & 120                                      & ~\cite{Arrieta2017,Arrieta2018}                  & 12                                    & \textcolor{black}{81.25\%   }                           & \textcolor{black}{89.29\%  }                            & \textcolor{black}{50\%  }                                                            \\ 
Two Tanks           & 498                                  & 11                                   & 7                                     & 150                                      & Random                   & 6                                     & \textcolor{black}{89.13\%   }                           & \textcolor{black}{91.67\%    }                          & \textcolor{black}{46.67\%  }                                                           \\ \hline
\end{tabular}
  \vspace{-0.55cm}

\end{table}

\subsection{Evaluation metric}

The Average Percentage of Faults Detected (APFD) was selected to evaluate the approach as it is the most used metric to evaluate test case prioritization approaches~\cite{Catal2013}. To measure the APFD, let T be a test suite containing $n$ test cases and F be a set of $m$ mutants detected by T. Let $TF_i$ be the index of the first test case in an ordering $ \pi $ of T that detects mutant $i$. Given an ordering $\pi$ of the test suite T, the APFD metric is defined as expressed in Equation \ref{eq:APFD}.

\begin{equation}
\label{eq:APFD}
APFD(\pi) = 1-\dfrac{\sum_{i=1}^{m}TF_i}{n\cdot m} + \dfrac{1}{2n}   
\end{equation}

Similar to the case of test cases, we used the same exact mutants as prior studies~\cite{Arrieta2019a,Arrieta2019} in order our findings to be comparable with those test case selection studies. To the best of our knowledge, the selected models are fault-free. However, employing mutation testing in the context of test prioritization has several advantages: it enables the systematic evaluation of test effectiveness, helps uncover subtle faults distributed across different parts of the code (or models, in this case), and provides a quantitative measure for comparing the fault detection capabilities of different test prioritization strategies. For the case of Swimairspeed, we generated and filtered out the mutants following the same criteria as Arrieta et al.~\cite{Arrieta2019a,Arrieta2019}.


\subsection{Statistical tests and algorithm runs}

For the selected techniques, it might be possible that at some points more than one option can be chosen by the test case prioritization algorithm because the selected test quality metric is the same in two or more test cases. As a result, tie-breaking strategies are required. As in previous studies~\cite{Henard2016}, we opted to randomly break these ties when they happen. To account for these random variations, we run each of the techniques 100 times and employed statistical tests to compare the different techniques, as proposed by Arcuri and Briand~\cite{Arcuri2011}. 

After obtaining the raw results by running the selected prioritization techniques, we employed the Shapiro-Wilk test to assess whether the data was normally distributed or not. Since the data was not normally distributed, we employed the Mann-Whitney U-test to obtain the statistical significance between two different techniques. The significance level was set to 5\%, meaning that there was a statistical significance if the p-value was lower than 0.05. The Vargha and Delaney \^{A}$_{12}$ value was also used to assess the difference existing between the techniques.

\section{Analysis of Results and Discussion}
\label{sec:results}
We now analyze the results and discuss their implications by answering the RQs. Table~\ref{tab:ResultsA12} reports the obtained statistical tests for the first three RQs. Specifically, the values reported in the table provide the Vargha and Delaney \^{A}$_{12}$ values, used to assess the effect size between two techniques. The \^{A}$_{12}$ measures, according to the obtained results, the probability that employing the technique in column \textit{Technique 1} will be more effective than the technique in column \textit{Technique 2}. A value higher than 0.5 means that the test case prioritization method in column \textit{Technique 1} is more effective than the method in column \textit{Technique 2}, whereas a value lower than 0.5 means the opposite.

\begin{sidewaystable}[!htbp]
	\centering
    \scriptsize
	\caption{RQ1. RQ2. RQ3: Results of the Vargha and Delaney \^{A}$_{12}$ values between two test prioritization techniques. Boldface values refer to statistical significance. i.e. p-value $\textless$ 0.05 for the Mann-Whitney U-test}
	\label{tab:ResultsA12}
\begin{tabular}{lllcccccc}
	\cline{2-9}
	& \textbf{Technique 1} & \textbf{Technique 2} & \textbf{\begin{tabular}[c]{@{}c@{}}Car \\ Window\end{tabular}} & \textbf{EMB}  & \textbf{\begin{tabular}[c]{@{}c@{}}Cruise \\ Controller\end{tabular}} & \textbf{Swimairspeed} & \textbf{\begin{tabular}[c]{@{}c@{}}AC \\ Engine\end{tabular}} & \textbf{\begin{tabular}[c]{@{}c@{}}Two \\ Tanks\end{tabular}} \\ \hline
	\multicolumn{1}{l}{\multirow{11}{*}{\textbf{RQ1}}} & AP-Ins          & Baseline        & \textbf{0.87}                                                  & \textbf{0.98} & 0.45                                                                  & \textbf{\textcolor{black}{1.00}}         & \textbf{1.00}                                                 & \textbf{0.90}                                                 \\ 
	\multicolumn{1}{l}{}                               & AP-Disc         & Baseline        & \textbf{1.00}                                                  & \textbf{0.96} & \textbf{0.31}                                                         & \textbf{\textcolor{black}{1.00}}         & \textbf{1.00}                                                 & \textbf{0.88}                                                 \\ 
	\multicolumn{1}{l}{}                               & AP-GTI          & Baseline        & \textbf{0.99}                                                  & \textbf{0.62} & \textbf{0.60}                                                         & \textbf{\textcolor{black}{1.00}}         & \textbf{1.00}                                                 & \textbf{0.97}                                                 \\ 
	\multicolumn{1}{l}{}                               & SB-IS           & Baseline        & \textbf{0.93}                                                  & \textbf{0.81} & \textbf{0.09}                                                         & \textbf{\textcolor{black}{1.00}}         & \textbf{1.00}                                                 & \textbf{0.40}                                                 \\ 
	\multicolumn{1}{l}{}                               & SB-OS           & Baseline        & \textbf{1.00}                                                  & \textbf{0.64} & \textbf{0.36}                                                         & \textbf{\textcolor{black}{1.00}}         & \textbf{0.99}                                                 & \textbf{0.32}                                                 \\ 
	\multicolumn{1}{l}{}                               & Add-DC           & Baseline        & \textbf{1.00}                                                  & \textbf{0.84} & 0.42                                                                  & \textbf{\textcolor{black}{1.00}}         & \textbf{0.99}                                                 & \textbf{0.97}                                                 \\ 
	\multicolumn{1}{l}{}                               & Add-CC           & Baseline        & \textbf{1.00}                                                  & \textbf{0.70} & \textbf{0.38}                                                         & \textbf{\textcolor{black}{1.00}}         & \textbf{0.99}                                                 & \textbf{0.97}                                                 \\ 
	\multicolumn{1}{l}{}                               & Add-MCDC         & Baseline        & \textbf{1.00}                                                  & \textbf{0.89} & \textbf{0.67}                                                         & \textbf{\textcolor{black}{1.00}}         & \textbf{0.99}                                                 & \textbf{0.97}                                                 \\ 
	\multicolumn{1}{l}{}                               & Tot-DC          & Baseline        & \textbf{1.00}                                                  & \textbf{0.99} & 0.46                                                                  & \textbf{\textcolor{black}{1.00}}         & \textbf{1.00}                                                 & \textbf{0.97}                                                 \\ 
	\multicolumn{1}{l}{}                               & Tot-CC          & Baseline        & \textbf{1.00}                                                  & \textbf{0.99} & \textbf{0.38}                                                         & \textbf{\textcolor{black}{1.00}}         & \textbf{1.00}                                                 & \textbf{0.97}                                                 \\ 
	\multicolumn{1}{l}{}                               & Tot-MCDC        & Baseline        & \textbf{1.00}                                                  & \textbf{0.99} & \textbf{0.35}                                                         & \textbf{\textcolor{black}{1.00}}         & \textbf{1.00}                                                 & \textbf{0.97}                                                 \\ \hline
	\multicolumn{1}{l}{\multirow{10}{*}{\textbf{RQ2}}} & AP-Ins          & AP-Disc         & \textbf{0.00}                                                  & \textbf{1.00} & \textbf{1.00}                                                         & \textcolor{black}{0.47}         & \textbf{1.00}                                                 & \textbf{1.00}                                                 \\ 
	\multicolumn{1}{l}{}                               & AP-Ins          & AP-GTI          & \textbf{0.00}                                                  & \textbf{1.00} & \textbf{0.00}                                                         & \textcolor{black}{0.47}       & \textbf{1.00}                                                 & \textbf{0.00}                                                 \\ 
	\multicolumn{1}{l}{}                               & AP-Ins          & SB-IS           & \textbf{0.23}                                                  & \textbf{0.96} & \textbf{0.97}                                                         & \textbf{\textcolor{black}{0.15}}         & \textbf{0.96}                                                 & \textbf{0.96}                                                 \\ 
	\multicolumn{1}{l}{}                               & AP-Ins          & SB-OS           & \textbf{0.00}                                                  & \textbf{1.00} & \textbf{0.75}                                                         & \textbf{\textcolor{black}{0.25}}         & \textbf{0.95}                                                 & \textbf{0.97}                                                 \\ 
	\multicolumn{1}{l}{}                               & AP-Disc         & AP-GTI          & \textbf{0.98}                                                  & \textbf{1.00} & \textbf{0.00}                                                         & \textcolor{black}{0.50}                  & \textbf{0.00}                                                 & \textbf{0.00}                                                 \\ 
	\multicolumn{1}{l}{}                               & AP-Disc         & SB-IS           & \textbf{1.00}                                                  & \textbf{0.93} & \textbf{0.92}                                                         & \textbf{\textcolor{black}{0.00}}         & \textbf{0.63}                                                 & \textbf{0.96}                                                 \\ 
	\multicolumn{1}{l}{}                               & AP-Disc         & SB-OS           & \textbf{0.88}                                                  & \textbf{1.00} & \textbf{0.32}                                                         & \textbf{\textcolor{black}{0.00}}         & \textbf{0.80}                                                 & \textbf{0.97}                                                 \\ 
	\multicolumn{1}{l}{}                               & AP-GTI          & SB-IS           & \textbf{0.83}                                                  & \textbf{0.11} & \textbf{0.98}                                                         & \textbf{\textcolor{black}{0.00}}         & \textbf{0.68}                                                 & \textbf{0.98}                                                 \\ 
	\multicolumn{1}{l}{}                               & AP-GTI          & SB-OS           & \textbf{0.12}                                                  & 0.43          & \textbf{0.97}                                                         & \textbf{\textcolor{black}{0.01}}         & \textbf{0.82}                                                 & \textbf{0.99}                                                 \\ 
	\multicolumn{1}{l}{}                               & SB-IS           & SB-OS           & \textbf{0.01}                                                  & \textbf{0.79} & \textbf{0.08}                                                         & \textbf{\textcolor{black}{0.76}}         & \textbf{0.71}                                                 & \textbf{0.58}                                                 \\ \hline
	\multicolumn{1}{l}{\multirow{18}{*}{\textbf{RQ3}}} & Add-DC           & Add-CC           & \textbf{0.75}                                                  & \textbf{0.67} & 0.54                                                                  & \textcolor{black}{0.47}                  & 0.51                                                          & 0.50                                                          \\ 
	\multicolumn{1}{l}{}                               & Add-DC           & Add-MCDC         & \textbf{0.84}                                                  & \textbf{0.40} & \textbf{0.23}                                                         & \textcolor{black}{\textbf{0.75}}                  & 0.52                                                          & 0.50                                                          \\ 
	\multicolumn{1}{l}{}                               & Add-DC           & Tot-DC          & 0.42                                                           & \textbf{0.23} & 0.45                                                                  & \textbf{\textcolor{black}{0.98}}         & \textbf{0.11}                                                 & 0.50                                                          \\ 
	\multicolumn{1}{l}{}                               & Add-DC           & Tot-CC          & \textbf{0.87}                                                  & \textbf{0.26} & 0.54                                                                  & \textbf{\textcolor{black}{0.98}}         & \textbf{0.12}                                                 & 0.50                                                          \\ 
	\multicolumn{1}{l}{}                               & Add-DC           & Tot-MCDC        & \textbf{0.93}                                                  & \textbf{0.23} & 0.57                                                                  & \textbf{\textcolor{black}{0.75}}                  & \textbf{0.11}                                                 & 0.50                                                          \\ 
	\multicolumn{1}{l}{}                               & Add-CC           & Add-MCDC         & \textbf{0.66}                                                  & \textbf{0.25} & \textbf{0.17}                                                         & \textcolor{black}{\textbf{0.76}}                  & 0.51                                                          & 0.50                                                          \\ 
	\multicolumn{1}{l}{}                               & Add-CC           & Tot-DC          & \textbf{0.17}                                                  & \textbf{0.08} & \textbf{0.39}                                                         & \textbf{\textcolor{black}{0.98}}         & \textbf{0.09}                                                 & 0.50                                                          \\ 
	\multicolumn{1}{l}{}                               & Add-CC           & Tot-CC          & 0.55                                                           & \textbf{0.09} & 0.50                                                                  & \textbf{\textcolor{black}{0.98}}       & \textbf{0.10}                                                 & 0.50                                                          \\ 
	\multicolumn{1}{l}{}                               & Add-CC           & Tot-MCDC        & \textbf{0.63}                                                  & \textbf{0.08} & 0.51                                                                  & \textbf{\textcolor{black}{0.76}}                  & \textbf{0.10}                                                 & 0.50                                                          \\ 
	\multicolumn{1}{l}{}                               & Add-MCDC         & Tot-DC          & \textbf{0.10}                                                  & \textbf{0.36} & \textbf{0.76}                                                         & \textbf{\textbf{\textcolor{black}{0.92}}}         & \textbf{0.11}                                                 & 0.50                                                          \\ 
	\multicolumn{1}{l}{}                               & Add-MCDC         & Tot-CC          & \textbf{0.34}                                                  & \textbf{0.39} & \textbf{0.83}                                                         & \textbf{\textbf{\textcolor{black}{0.92}}}         & \textbf{0.12}                                                 & 0.50                                                          \\ 
	\multicolumn{1}{l}{}                               & Add-MCDC         & Tot-MCDC        & \textbf{0.39}                                                  & \textbf{0.36} & \textbf{0.90}                                                         & \textcolor{black}{0.47}                 & \textbf{0.11}                                                 & 0.50                                                          \\ 
	\multicolumn{1}{l}{}                               & Tot-DC          & Tot-CC          & \textbf{1.00}                                                  & \textbf{0.60} & \textbf{0.61}                                                         & \textcolor{black}{0.50}                  & 0.53                                                          & 0.50                                                          \\ 
	\multicolumn{1}{l}{}                               & Tot-DC          & Tot-MCDC        & \textbf{1.00}                                                  & 0.50          & \textbf{0.65}                                                         & \textbf{\textcolor{black}{0.06}}         & 0.51                                                          & 0.50                                                          \\ 
	\multicolumn{1}{l}{}                               & Tot-CC          & Tot-MCDC        & \textbf{0.90}                                                  & \textbf{0.40} & 0.51                                                                  & \textbf{\textcolor{black}{0.06}}         & 0.48                                                          & 0.50                                                          \\ \hline
\end{tabular}
\end{sidewaystable}

\begin{figure}[h]
	
\begin{subfigure}{.3\textwidth}
		\centering
		\includegraphics[trim=100 260 100 270, clip, width=1\linewidth]{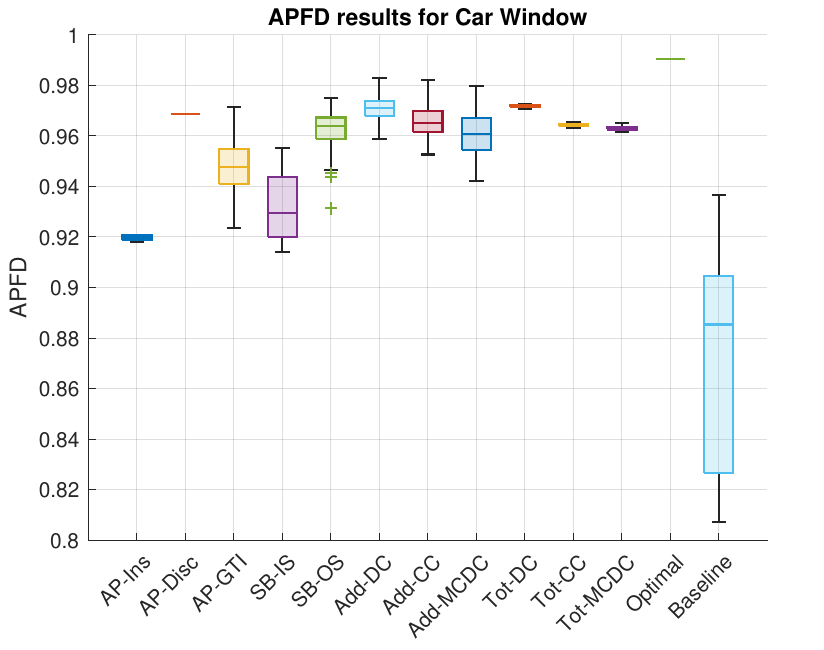}
	\end{subfigure}
	\begin{subfigure}[t]{.3\textwidth}
		\centering
		\includegraphics[trim=100 260 100 270, clip, width=1\linewidth]{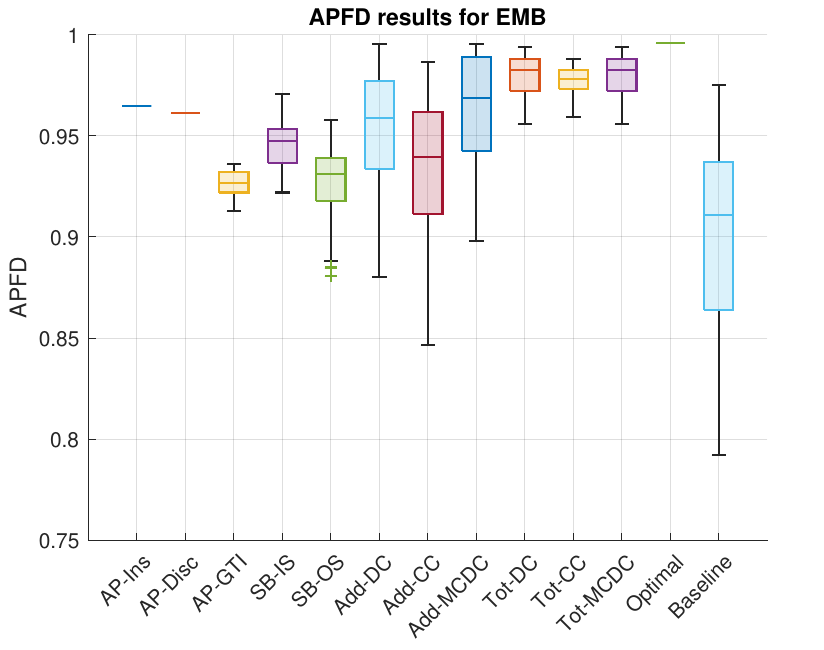}
	\end{subfigure}
	\begin{subfigure}[t]{.3\textwidth}
		\centering
		\includegraphics[trim=100 260 100 270, clip, width=1\linewidth]{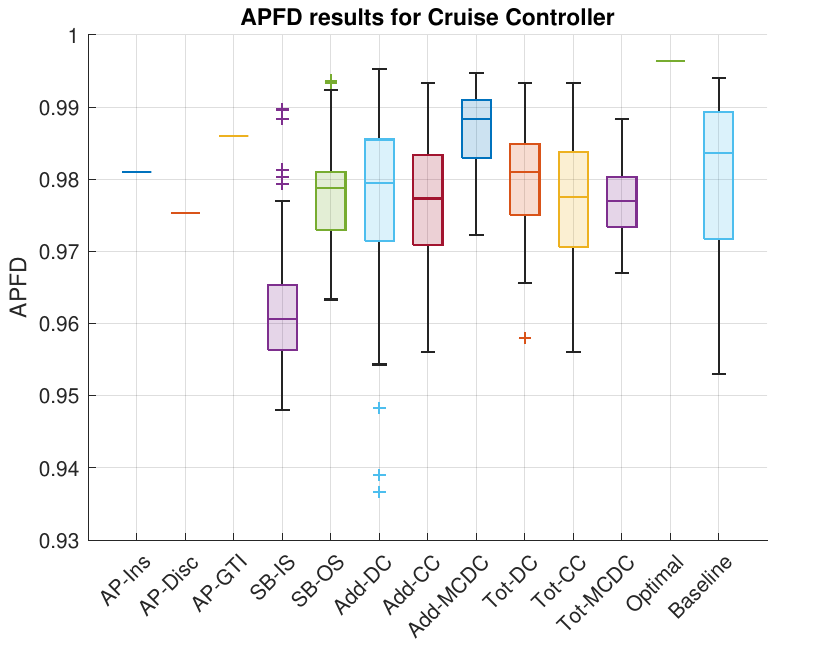}
	\end{subfigure}
	\begin{subfigure}[t]{.3\textwidth}
		\centering
		\includegraphics[trim=100 260 100 270, clip, width=1\linewidth]{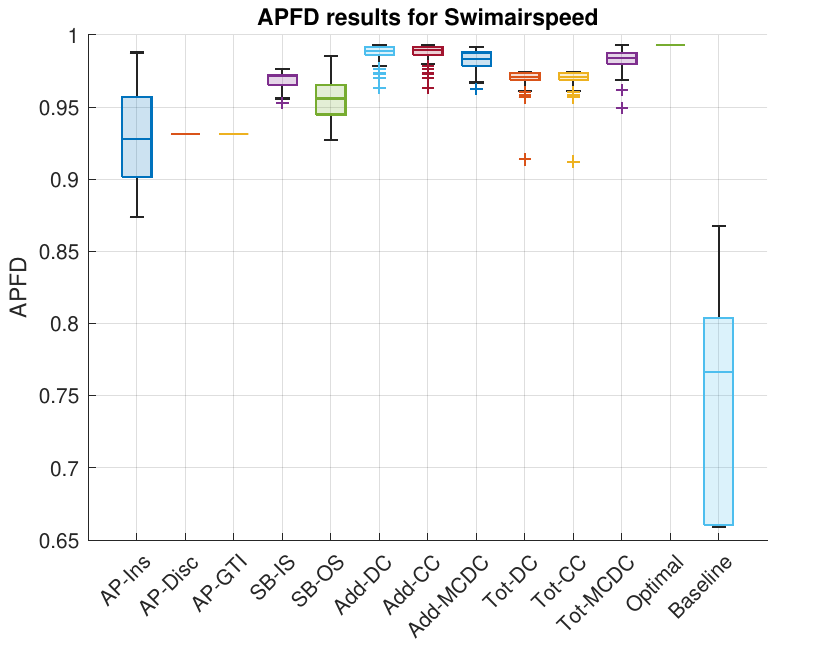}
	\end{subfigure}
	\begin{subfigure}[t]{.3\textwidth}
		\centering
		\includegraphics[trim=100 260 100 270, clip, width=1\linewidth]{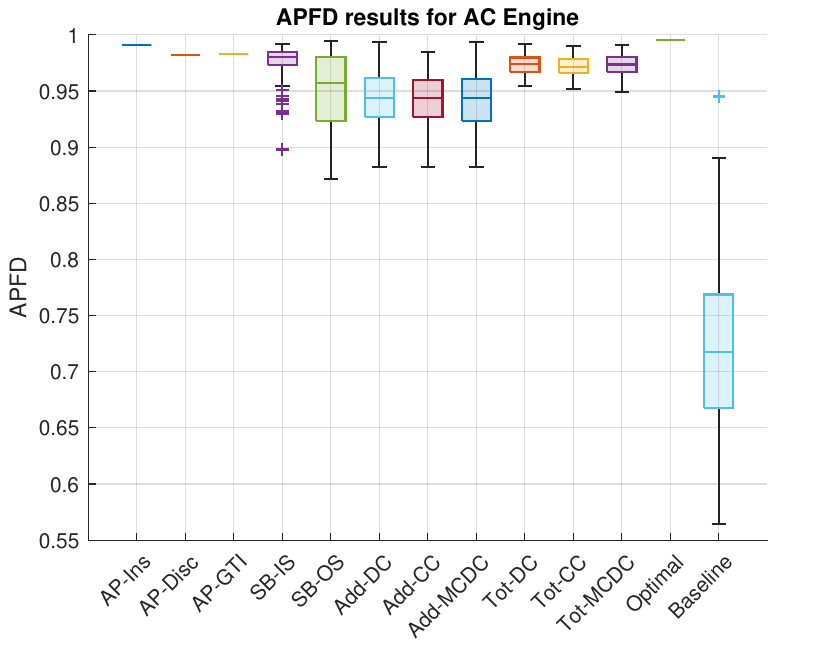}
	\end{subfigure}
	\begin{subfigure}[t]{.3\textwidth}
		\centering
		\includegraphics[trim=100 260 100 270, clip, width=1\linewidth]{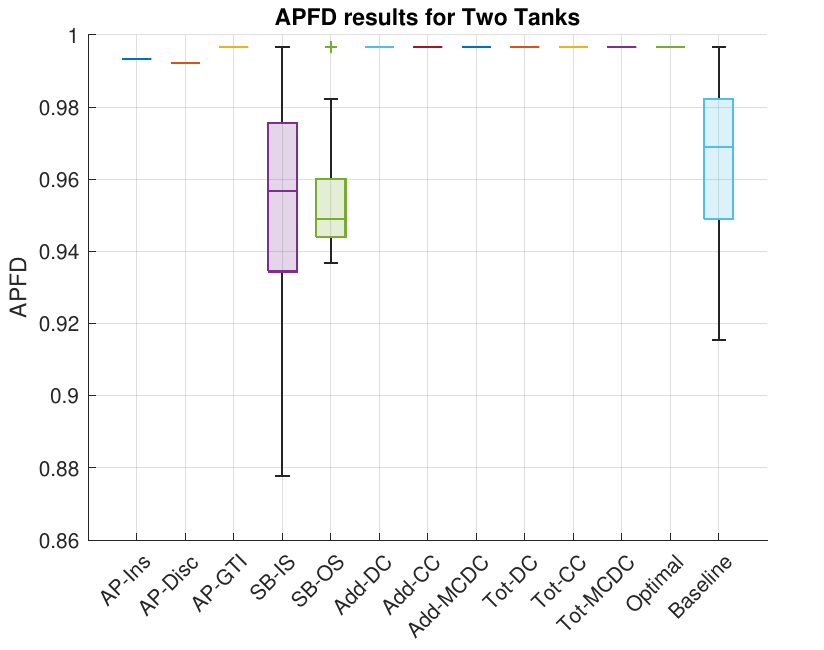}
	\end{subfigure}
	\caption{Distribution of the obtained APFD values for the six subject models and the selected techniques}
	\label{fig:APFDResults}
\end{figure}

\subsection{RQ1 -- Sanity Check}

As shown in Table~\ref{tab:ResultsA12} and Figure~\ref{fig:APFDResults}, in all subject models there was more than one technique that significantly outperformed the selected baseline test prioritization technique. In fact, with the exception of the Cruise Controller and the Two Tanks subject models, all the techniques outperformed the baseline test prioritization technique with statistical significance. For the Cruise Controller model, two techniques (AP-GTI and Add-MCDC) outperformed the baseline algorithm with statistical significance. Conversely, the baseline algorithm outperformed eight of the selected test prioritization techniques with statistical significance. A reason why the baseline algorithm outperformed some of the techniques could be the simplicity of this subject model, which is the smallest one from the six. This hypothesis matches with one of the lessons learned from Quo et al.~\cite{Luo2019}, where they found that test prioritization performs better on larger programs. As for the Two Tanks subject models, all the techniques with the exception of SB-IS and SB-OS techniques outperformed with statistical significance the baseline technique. Nevertheless, for the remaining subject models, all the selected test prioritization techniques outperformed the selected baseline algorithm with statistical significance, as can be seen in Table \ref{tab:ResultsA12}. This means that the selected test prioritization techniques along with their test quality metric defined with respect to the six Simulink models considered seem to be effective. We can thus answer the first RQ as follows:

\begin{custombox}{RQ1}
\textit{Overall, the selected techniques are better than the selected baseline test prioritization algorithm, meaning that they are effective in this context and can be recommended for practitioners.}
\end{custombox}

\subsection{RQ2 -- Black-box techniques}

When comparing the performance of black-box test prioritization techniques, it can be seen that for all subject models except for Swimairspeed, the test prioritization techniques that were based on anti-patterns stand out over the test prioritization techniques based on test similarity metrics. These results are consistent with those obtained by Arrieta et al., for the context of test selection approaches~\cite{Arrieta2019a}. %

Similar to a study conducted in the context of test selection~\cite{Arrieta2019a,Arrieta2018a,arrieta2023some}, anti-patterns seem to be, in general, more appropriate test quality metrics than similarity-based metrics. However, similarity-based techniques outperformed them in the Swimairspeed case study system. A potential reason could be the types of outputs the Swimairspeed system provides, which are related to discrete (software) values providing warning flags. In contrast, the rest of the case study systems provide outputs that correspond to time-continuous physical phenomena (e.g., position of the 4 windows in a car in CW, or the engine speed (in rpm-s) in the ACEngine system).  

By analysing the characteristics of each case study from Table~\ref{tab:CaseStudy}, we could not find one specific pattern by which we could recommend a test quality metric over the others for test prioritization. We conclude that the best anti-pattern metric for each case may depend on the system and type of test cases that are used, and thus, it is difficult to recommend one such metric. All anti-patterns based metrics showed strong APFD values, all having a median above 0.9 in all case study systems. Specifically, we see that AP-Disc performs consistently well, showing median APFD values above 0.95 in 5 out of the 6 case study systems, and not having any single APFD value for all runs below 0.90. 
Notably, the variances of the APFD results with these subject models for the anti-patterns-based test prioritization techniques were minimal. This is because the anti-pattern degree for each test case is unique in most cases and, therefore, ties usually do not exist. As for the similarity-based metrics, similar to previous works in the context of simulation-based testing~\cite{Arrieta2019a,Arrieta2018a}, results were not as good as the anti-patterns. In two out of the six case study systems, the results were even worse than the baseline. Only in one case study systems (i.e., Swimairspeed), similarity-based techniques performed better than anti-pattern ones. We further discuss the key issues and potential solutions for similarity-based techniques in Section~\ref{sec:qa}.

Having discussed this, we can answer the second RQ as follows:

\begin{custombox}{RQ2}	
\textit{Anti-patterns based techniques are more appropriate than techniques based on test case similarity when the outputs of the models yield physical values
Similarity-based techniques performed better in Swimairspeed, which had discrete outputs related to software control values (e.g., flags). }
\end{custombox}

\subsection{RQ3 -- White-box techniques}

The third RQ aimed at assessing the performance of six white-box test prioritization techniques. In the two tanks subject model, all techniques performed equally in terms of the APFD. When considering statistical tests, in three simulation models (Car Window, EMB and ACEngine), Tot-DC outperformed the remaining techniques, whereas in the remaining two (i.e., Cruise Controller and Swimairspeed), the Add-MCDC and Add-CC technique performed best. The good performance of the total techniques might be because a single test case covering a test objective might not be enough to detect a fault. This finding also explains the poor performance of white-box techniques in the context of multi-objective test case selection of simulation models~\cite{Arrieta2019a}. Furthermore, the subject models where Tot-DC performed best were the largest in terms of number of blocks. In larger models, more test cases are required to achieve a larger additional coverage, whereas total techniques (e.g., Tot-DC) consistently prioritize test cases that cover large portions of objectives. In the presence of complex faults, some test cases may exercise the fault without triggering a failure. In such cases, the Tot-DC may have an advantage over others because it prioritizes test cases that maximize the overall decision coverage rather than just incremental gains. This ensures that test cases selected early in the prioritization process exercise a wide variety of decision points, potentially exposing faults that only manifest under interactions across multiple decisions. In contrast, additional greedy techniques might focus on test cases that cover only a single new test objective. Another explanation could be that some faults may require the interaction of multiple decisions, which might not be covered by additional greedy techniques that solely adds minimal new coverage. Instead, a test case focusing on large total decision coverage objectives will simultaneously exercise various parts of the model, increasing the chances of revealing faults that are dependent on various the interaction of multiple decision points of the model. 
Moreover, similar to what happened in the previous RQ with anti-pattern metrics, the variance of the APFD results with the total test prioritization approaches seems to be lower than the other techniques. Low variance in results is a positive aspect for using this technique in practice. 
Thus, we can answer the third RQ as follows:

\begin{custombox}{RQ3}	
	\textit{Overall, when considering median APFD values, all white-box techniques show high competitiveness at prioritizing test cases in the context of Simulink models. However when considering statistical significance, we found that Tot-DC performed best in three out of six subject models, and obtained the same results as the rest in a fourth subject models; all these models were the largest ones in terms of number of blocks. In contrast, Add-MCDC and Add-CC performed best in the two smallest models (i.e., Cruise Controller and Swimairspeed).}
\end{custombox}

\subsection{RQ4 -- Overall best technique}

RQ4 aimed at selecting the best overall metric. To this end, we compared the best techniques obtained from RQ2 and RQ3 for each subject model. As can be seen in Table \ref{tab:RQ4Tests}, for the first four subject models (i.e., CW, EMB, CC and Swimairspeed), the best white-box techniques outperformed the black-box techniques with statistical significance. Conversely, for the ACEngine subject models, the best black-box technique outperformed the white-box onewith statistical significance. Lastly, for the Two Tanks subject model, the \^{A}$_{12}$ value was 0.5, meaning that both techniques performed equally.

\begin{table}
	\scriptsize
	\centering
	\caption{RQ4: Results of the Vargha and Delaney \^{A}$_{12}$ values between the best two test prioritization techniques obtained from RQ2 and RQ3}
	\label{tab:RQ4Tests}
	\begin{tabular}{lllllc}
		\hline
		\multicolumn{1}{c}{\textbf{\begin{tabular}[c]{@{}c@{}}Best BB\\ Technique\end{tabular}}} & \multicolumn{1}{c}{\textbf{\begin{tabular}[c]{@{}c@{}}Best WB \\ Technique\end{tabular}}} & \multicolumn{1}{c}{\textbf{Subject}} & \multicolumn{1}{c}{\textbf{\^{A}$_{12}$}} & \multicolumn{1}{c}{\textbf{p-value}} & \textbf{Overal Best} \\ \hline
		AP-Disc                                                                                    & Tot-DC                                                                                     & Car Window                                    & 0                                 & \textless{}0.0001                     & Tot-DC               \\ 
		AP-Ins                                                                                     & Tot-DC                                                                                     & EMB                                   & 0.11                              & \textless{}0.0001                     & Tot-DC               \\ 
		AP-GTI                                                                                     & Add-MCDC                                                                                    & Cruise Controller                                    & 0.36                              & \textless{}0.0001                     & Add-MCDC              \\ 
		SB-IS                                                                                     & Add-CC                                                                                      & Swimairspeed                          & 0.0125                                 & \textless{}0.0001                     & Add-CC               \\ 
		AP-Ins                                                                                    & Tot-DC                                                                                     & ACEngine                              & 0.99                           & \textless{}0.0001                     & AP-Ins             \\ 
		AP-GTI                                                                                     & ALL                                                                                        & TwoTanks                              & 0.5                               & -                                     & same                 \\ \hline
	\end{tabular}
\end{table}

Unlike in the context of multi-objective test case selection~\cite{Arrieta2019a}, where black-box approaches significantly outperformed white-box ones, for the context of test prioritization, white-box techniques are as competitive as black-box ones, performing better in most of the studied case study systems. 
A reason why white-box metrics performed well in the context of test prioritization and not that well in the context of test selection could be that, in the latter case, the objective was to reduce the test execution time as much as possible while incrementing test coverage. Because of this, such techniques will tend to execute specific test objectives only once in order to maximize the coverage while minimizing execution times, which might not be a good approach for detecting (all the) faults. Regardless, it is still possible for coverage-based test prioritization techniques to execute (the same) test objectives multiple times, especially for total approaches. Although when considering statistical results, in general, white-box techniques were better than the black-box ones, when considering median values, the black-box techniques were not far from the white-box ones. Moreover, in one of the models (i.e., ACEngine) black-box techniques showed better performance, 
and obtained the same APFD scores as white-box ones in another one. Furthermore, for those subject models where white-box techniques performed better, the median values of the APFD for the black-box techniques were quite close. Besides, the variances of white-box techniques were larger than the variances of the black-box anti-patterns based test prioritization techniques, which means that tie-breaking is required more often for the former.

We further compared the overall best technique extracted with the optimal one, which was obtained by considering the information of test cases exposing the seeded faults. As can be seen in Table~\ref{tab:RQ5}, the best techniques are quite close to the optimal one. In fact, for one of the subject simulation models (i.e., Two Tanks), the best techniques yielded optimal results. For the remaining subject models, the optimal test order outperformed the best technique by a margin of 0.4\% to 1.9\% in terms of the APFD. We consider these values fairly good, which means that the proposed test prioritization algorithms can be useful in practice and close to optimality. In conclusion, the fourth RQ can be answered as follows:

\begin{custombox}{RQ4}	
	\textit{In four of the models, the best white-box techniques performed better than the black-box ones. The best black-box technique only performed better than the white-box one in one of the models. However, when considering average values, the differences were quite small. 
    The best test prioritization techniques are quite close to the optimal test ordering in all case studies, which means that they can be recommended for practitioners.}
\end{custombox}





\subsection{RQ5 -- Test prioritization time}
\label{sec:time}

RQ5 aimed at evaluating the test prioritization time required by each technique. The overall results are reported in Table \ref{tab:ResultsTime}, where the average running time for each technique and its standard deviation is shown. From these results, it can be seen that black-box metrics were, in the cases considered, significantly better than white-box techniques in this aspect; this was corroborated by statistical tests, where black-box techniques were shown to outperform all white-box ones with statistical significance for the six subject models (in all cases). When taking a closer look at black-box test prioritization techniques, it can be seen that the techniques based on anti-patterns, which require less than a millisecond on average, are faster than the techniques based on similarity. Nevertheless, both similarity-based metrics are around 0.01 seconds on average, which is something that is, by far, affordable by practitioners.

\begin{table}
\scriptsize
\centering
\caption{Improvement extent by the optimal test prioritization with respect to the best technique for each subject model}
\label{tab:RQ5}
\begin{tabular}{llr}
\hline
\textbf{Best technique} & \textbf{\begin{tabular}[c]{@{}l@{}}Subject\\ model\end{tabular}} & \multicolumn{1}{l}{\textbf{\begin{tabular}[c]{@{}l@{}}Improvement\\ Extent\end{tabular}}} \\ \hline
Tot-DC                  & CW                                                               & 1.9\%                                                                                      \\ 
Tot-DC                  & EMB                                                              & 1.63\%                                                                                     \\ 
Add-MCDC                 & CC                                                               & 0.92\%                                                                                     \\ 
Add-CC                 & Swimairspeed                                                     & 0.53\%                                                                                        \\ 
AP-Inst                 & ACEngine                                                         & 0.42\%                                                                                     \\ 
same                    & TwoTanks                                                         & 0\%                                                                                        \\ \hline
\end{tabular}
\end{table}

 As for white-box metrics, it can be observed that additional techniques were much slower than total techniques. On average, total techniques required below 0.4 seconds, whereas additional techniques took more than 135 seconds for all subject models, and more than 180 seconds for two of the subject models. In order to get higher insights regarding the execution time of these algorithms, we launched a profiler that measured the time it takes for each line of code in the algorithm to execute.\footnote{We launched this with the Add-DC technique along with the two tanks subject model, but it is expectable that similar results will yield with other additional greedy-based techniques and other subject models too} When we saw the results, we found out that the bottleneck was on the API calls for measuring the coverage of Simulink models. In fact, for the selected technique, five lines of code from the algorithm took 99.8\% of the running time. These five lines of code are all related with the measurement of the white-box code coverage with the API. We can therefore answer the last RQ as follows:

\begin{custombox}{RQ5}
	\textit{Black-box test prioritization techniques are significantly faster than white-box ones. On the other hand, total techniques are much faster than additional techniques. The overhead of the additional techniques is mainly caused by the high number of calls to the coverage API, which significantly increases the test prioritization time.}
\end{custombox}



\subsection{Discussion}

\subsubsection{Concluding Remark and Recommendations}
\label{sec:CR}

The empirical evaluation showed that there is no single best technique for prioritizing test cases for Simulink models. Nevertheless, there are some insights that are worth mentioning as a summary:

\begin{itemize}[leftmargin=10pt]
	\item White-box techniques are overall slightly stronger than black-box ones, which contrasts with previous studies \cite{Arrieta2019a}.
    \item Among the black-box techniques, overall, anti-patterns based test prioritization techniques performed better than distance-based techniques. 
	\item Among the white-box techniques, the total technique along with the DC metric (Tot-DC) performed best in three out of six subject models, and it performed equally in the two tanks subject model. This technique was the best for large subject models, whereas Add-MCDC and Add-CC were the best for smaller ones.
	\item Black-box techniques were 
	faster than white-box ones. Nevertheless, total greedy based test prioritization techniques are fast enough to be applied in practice. 
	\item Similarity-based test prioritization techniques were 
	overall less competitive than the remaining techniques. In Section \ref{sec:qa} we give some insights of why these metrics may fail to perform well in the context of Simulink models and guide future researchers on potential solutions. 
    \item Results may differ depending on the Simulink model at which testing is being applied. However, our study shows evidence that either AP-Ins or Tot-DC could perform competitively, and thus, 
we recommend either of both when test prioritization methods need to be applied.
\end{itemize}

\begin{table}[t!]

	\scriptsize
	\centering
	\caption{RQ5: Average running time (in seconds) and standard deviation for each test prioritization technique}
	\label{tab:ResultsTime}
	\resizebox{\textwidth}{!}{
		\begin{tabular}{llllllllllllll}
			\cline{3-14}
			&        & \textbf{AP-ins}  & \textbf{AP-Disc} & \textbf{AP-GTI}  & \textbf{SB-IS}   & \textbf{SB-OS}   & \textbf{AR-DC}     & \textbf{AR-CC}     & \textbf{AR-MCDC}      & \textbf{Tot-DC}  & \textbf{Tot-CC}  & T\textbf{ot-MCDC} & \textbf{Baseline} \\ \hline
			\multicolumn{1}{l}{\multirow{2}{*}{\textbf{Car Window}}}        & Avg    & 0.00071 & 0.00053 & 0.00040 & 0.01474 & 0.01295 & 152.39616 & 152.09648 & 152.26632 &  0.34866 & 0.34939 & 0.34914  & 0.01410  \\ 
			\multicolumn{1}{l}{}                                   & stdDev & 0.00425 & 0.00279 & 0.00187 & 0.01398 & 0.00458 & 4.08801   & 3.79466   & 3.56654   & 0.01348 & 0.01592 & 0.01365  & 0.00993  \\ \hline
			\multicolumn{1}{l}{\multirow{2}{*}{\textbf{EMB}}}               & Avg    & 0.00032 & 0.00024 & 0.00023 & 0.01508 & 0.01465 & 156.87667 & 151.42402 & 156.90839 &  0.38903 & 0.38674 & 0.38819  & 0.01485  \\ 
			\multicolumn{1}{l}{}                                   & stdDev & 0.00029 & 0.00016 & 0.00014 & 0.00187 & 0.00137 & 3.79300   & 3.56347   & 4.03221   & 0.02409 & 0.02171 & 0.02260  & 0.00257  \\ \hline
			\multicolumn{1}{l}{\multirow{2}{*}{\textbf{Cruise Controller}}} & Avg    & 0.00087 & 0.00028 & 0.00025 & 0.01501 & 0.01340 & 146.45570 & 51.91876  & 107.30989 &  0.28479 & 0.28276 & 0.28205  & 0.01417  \\ 
			\multicolumn{1}{l}{}                                   & stdDev & 0.00569 & 0.00063 & 0.00024 & 0.00732 & 0.00036 & 3.42112   & 0.91751   & 3.55128   &  0.00557 & 0.00617 & 0.00421  & 0.00683  \\ \hline
		\multicolumn{1}{l}{\multirow{2}{*}{\textbf{Swimairspeed}}}      & Avg    & 0.00084 & 0.00041 & 0.00037 & 0.01805 & 0.01593 & 197.84829 & 194.58293 & 187.34234 &  0.30118 & 0.3102 & 0.31038  & 0.01391  \\ 
			\multicolumn{1}{l}{}                                   & stdDev & 0.000314 & 0.00026 & 0.00033 & 0.00956 & 0.00079 & 10.53843   & 10.3582   & 10.16534     & 0.00476 & 0.00493 & 0.00475  & 0.00523  \\ \hline
			\multicolumn{1}{l}{\multirow{2}{*}{\textbf{AC Engine}}}         & Avg    & 0.00615 & 0.00032 & 0.00028 & 0.01407 & 0.01220 & 136.40267 & 135.96562 & 139.81619  & 0.42410 & 0.42207 & 0.42517  & 0.01343  \\ 
			\multicolumn{1}{l}{}                                   & stdDev & 0.04902 & 0.00060 & 0.00028 & 0.00718 & 0.00212 & 16.41926  & 16.00144  & 39.72185  &  0.05656 & 0.05402 & 0.05721  & 0.01229  \\ \hline
			\multicolumn{1}{l}{\multirow{2}{*}{\textbf{Two Tanks}}}         & Avg    & 0.00039 & 0.00024 & 0.00022 & 0.01589 & 0.01367 & 184.91517 & 184.94793 & 180.45511  & 0.30852 & 0.30580 & 0.30632  & 0.01408  \\ 
			\multicolumn{1}{l}{}                                   & stdDev & 0.00084 & 0.00018 & 0.00017 & 0.00956 & 0.00041 & 9.76272   & 9.79499   & 9.73668     & 0.00476 & 0.00342 & 0.00405  & 0.00530  \\ \hline
	\end{tabular}}
\end{table}

Based on our experiments and observations, we offer the following recommendations for practitioners using test prioritization techniques with Simulink models:

\begin{itemize}[leftmargin=10pt]

\item \textbf{Recommendation 1:} There is no clear winner among the selected techniques. We believe that results may vary depending on the system types as well as selected test cases. Since test prioritization of CPSs modeled in Simulink is foreseen to be applied both at different test levels (i.e., MiL, SiL, HiL) as well as a regression testing context, we recommend practitioners to trace the correlation between different anti-patterns and coverage metrics with respect to found faults. However, this information is usually not available, for which we recommend practitioners to further analyze our following recommendations.

\item \textbf{Recommendation 2:} In general, we do not recommend the use of similarity-based test prioritization techniques. We found that the Euclidean distance may fail to capture some important signal features in the context of Simulink models testing (further discussed in Section~\ref{sec:qa}).

\item \textbf{Recommendation 3:} Overall, white-box techniques perform competitively. Thus, we recommend their usage in the context of Simulink models for test prioritization. We recommend total techniques, and especially Tod-DC, when the Simulink model is relatively large, both, due to their high effectiveness and efficiency. In contrast, if the model is small, we recommend the Add-MCDC or Add-CC, although if the running time is an issue it can eventually become a problem.

\item \textbf{Recommendation 4:} White-box techniques are not always available (e.g., if the subsystem is from an external company that does not provide the model, but only the executable). In those cases, we recommend practitioners to resort to anti-patterns based test prioritization techniques, but only when the outputs of the models relate to signals of physical phenomena (e.g., speed of a car). In contrast, if the outputs are related to discrete software-related values (e.g., flags, communication signals), although not ideal, we recommend the use SB-IS.

\end{itemize}

\subsubsection{Qualitative Analysis and Future Research Avenues}
\label{sec:qa}

\begin{figure}{ht}
    \centering
    \includegraphics[trim= 120 280 120 280,clip, width=\linewidth]{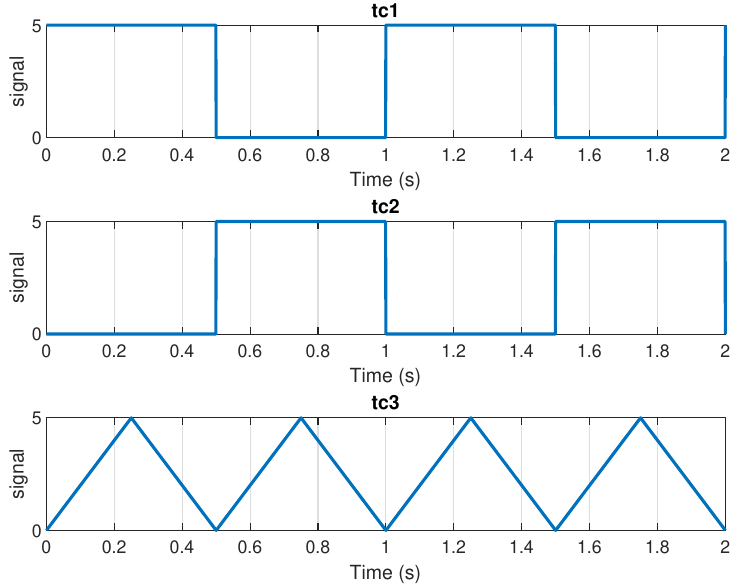}
    \caption{An example showing the problems of the Euclidean distance at capturing signal shapes}
    \label{fig:problem}
\end{figure}

This paper conducts the first attempt to assess traditional white-box and black-box techniques in the context of Simulink models test prioritization, with interesting insights discussed through 5 different RQs and summarized in Section~\ref{sec:CR}. By extracting such insights, we now conduct a qualitative analysis and extract future research avenues that can help guide researchers in their future research efforts:

\textit{\textbf{Research novel metrics for similarity-based test prioritization:}} Our results reveal that traditional similarity-based test prioritization methods are less competitive than other techniques. We believe that future research avenues could focus on researching novel distance metrics that consider signal shapes and other features that the Euclidean distance (i.e., the one employed in this and other studies~\cite{Matinnejad2015b,Matinnejad2016,Arrieta2018a,Arrieta2019a}) fails to capture. As an illustration of this problem, consider Figure \ref{fig:problem}, which includes 3 different test cases. Test cases $tc_1$ and $tc_2$ have very similar shapes (square signal with the same frequency), but the Euclidean distance between both of these test cases is the maximum (assuming no other signal exists in the test case). In many Simulink models, both of these test cases will test similar functionalities. In contrast, $tc_3$ provides a triangle wave signal. The Euclidean distance between this signal with $tc_1$ is lower than that with $tc_3$, whereas there is a high likelihood that this test case tests other properties not covered by $tc_1$ and $tc_2$. This example highlights a key limitation of relying solely on Euclidean distance as a similarity metric: it fails to account for the structural differences in signal shapes that can be critical for assessing the diversity and functional coverage of test cases. Future work could explore the development of more sophisticated distance metrics or embedding techniques that capture features such as signal shape, frequency content, and temporal patterns, enabling more effective prioritization strategies for simulation-based testing of CPSs in Simulink and other domains. 

\textit{\textbf{Criticality of bugs:}} The criticality of bugs in CPSs varies based on the violated requirements. For example, in autonomous vehicles, drivability requirements~\cite{formica2023test} related to comfort are less critical than safety requirements, such as collision avoidance. Bugs violating safety requirements are more crucial to detect early. Since our test prioritization techniques make no assumptions about which system properties are critical, comparing the types of bugs detected would be unfair. Future work could explore prioritization techniques that consider system criticality. For instance, a Domain Specific Language could be proposed to specify the criticality of Simulink subsystems or test outputs affecting safety. This information could later be processed by the prioritization technique. This would also require adapting the APFD metric to account for bug criticality.

\textit{\textbf{Severity of failures:}} Likewise, severity of bugs is also an aspect that could be measured in the future. Following with an example from the automotive domain, crashing a vehicle at \textit{20 km/h} is less critical than doing it at \textit{120 km/h}. In the future, it would be interesting to adapt the APFD metric to consider such aspect.

\textit{\textbf{Test prioritization for flaky tests:}} While the models in our benchmark are deterministic, some CPS simulators exhibit flakiness~\cite{amini2024evaluating}, meaning that tests may pass or fail inconsistently even when the system and simulator configurations remain unchanged. Moreover, in later stages (e.g., HiL), where hardware devices are involved, stochastic simulations are a common challenge. Adapting test prioritization techniques to address this issue presents a promising direction for future research.

\textit{\textbf{Adequacy techniques in AI-enabled CPSs:}} AI models (e.g., neural networks) are increasingly prevalent in the control of CPSs. However, traditional white-box techniques cannot be directly applied to assess the quality of test cases for these AI components. Exploring the integration of adequacy techniques from this domain (e.g., neuron coverage) with conventional white-box test prioritization techniques and evaluating their impact on fault detection rates represents another promising and unexplored research avenue.

\textit{\textbf{Tie-breaking strategies:}} As shown in Figure~\ref{fig:APFDResults}, many techniques have a large distribution. This distribution could eventually be reduced, and the APFD probably increased, by integrating tie-breaking strategies in the test prioritization techniques.


\section{Threats to Validity}
\label{sec:threats}

We now discuss the threats to validity of our study and how we tried to mitigate them. A potential \textbf{internal validity} threat of our empirical evaluation is related to the generated mutants. In the context of Simulink, models often have a physical layer that are typically modeled through complex mathematical equations. As a result, the execution of test cases is time consuming, making the use of a large set of mutants impractical. However, the amount of mutants used in this study is similar to other studies were MATLAB/Simulink models were used~\cite{Matinnejad2016,Arrieta2016a,Arrieta2016b,Arrieta2019,Matinnejad2015b,Liu2017,Liu2018,Liu2016,LeThiMyHanh2014,Hanh2016,Matinnejad2019}. Besides, we have tried to mitigate this threat by removing duplicated mutants, as recommended by Papadakis et al.~\cite{Papadakis2015}. An \textbf{external validity} threat in all software engineering empirical evaluations is related to the generalization of results. We used six subject models, which might not be enough to generalize our results. Nevertheless, we employed simulation models of different characteristics and sizes to reduce this threat. Additionally, one of the subject models, the EMB, was an industrial case study developed by Bosch. When referring to the size of the subject models, according to a previous paper where 391 public Simulink models were analyzed, more than half of the analyzed models had less than 100 blocks and around 75\% of the models had less than 300 blocks \cite{Chowdhury2018}. In our empirical study, five out of six subject models had from 130 to 498 blocks, meaning that their complexities in terms of model size are higher than most of the public subject models. In addition, four of the blocks from the Car Window subject model were state machines with 5 states and 21 transitions each. A \textbf{conclusion validity} threat in our study might be related to the stochastic nature of our experiments. This is caused by our strategy of randomly breaking ties between multiple test cases with the same adequacy score, which has been proposed in other studies (e.g.,~\cite{Henard2016}). To reduce the threat produced by the randomization of our algorithms, we ran each algorithm 100 times and applied statistical tests to analyze the results, as recommended by Arcuri and Briand~\cite{Arcuri2011}.

\section{Related Work}
\label{sec:relatedWork}

Testing has been widely applied to MATLAB/Simulink, the defacto tool for modeling and simulating CPSs. A number of studies have proposed automated test generation approaches, including mutation-based test generation~\cite{Hanh2016}, generation based on formal methods~\cite{He2011} and model checking~\cite{Mohalik2014}, search-based test case generation~\cite{Matinnejad2016,Matinnejad2017,Matinnejad2019,formica2022simulation,formica2022search}, falsification-based test generation~\cite{Menghi2020,nejati2023reflections,Formica2023} and fuzzing~\cite{su2024test}. Besides test generation, other approaches have proposed other testing activities, including automated fault localization~\cite{Liu2018,Liu2016,Liu2017}, test oracle generation~\cite{Menghi2019}, mutation testing~\cite{Zhan2005} and test case selection~\cite{Arrieta2018a,Arrieta2019a,Arrieta2016b,arrieta2023some,arrieta2023novel}. Unlike all these studies, this paper focuses on test case prioritization for regression testing of Simulink models. The use of anti-patterns for Simulink models was first proposed by Matinnejad et al.~\cite{Matinnejad2016,Matinnejad2017}. In this paper, we adapted the proposed anti-patterns measures to the context of test case prioritization. However, it is important to note that the definition of some measures for test suite generation has not prevented successfully reusing the same measures in test case prioritization in other contexts~\cite{Henard2016}.

As for test case prioritization in the context of Simulink models, Arrieta et al.~\cite{Arrieta2019,Arrieta2016a} considered information related to a historical database to prioritize test cases in a cost-effective manner by using weighted search algorithms. Matinnejad et al.~\cite{Matinnejad2019}, along with their test generation approach, proposed a complementary test case prioritization algorithm that uses a combination of model coverage information (white-box) and the output diversity of the test suites (black-box) as surrogate criteria for a greedy algorithm to estimate the probability of revealing new faults with each test case. In their mutation-based experiments with two industrial Simulink models, they conclude that this approach significantly outperforms random prioritization, as well as total and additional coverage-based algorithms (white box). The study we present differs from those ones by (1) focusing exclusively on either white-box or black-box techniques, (2) using traditionally employed test case prioritization approaches in the context of Simulink models and (3) providing an empirical evaluation for a total of 11 test case prioritization techniques.

Empirical evaluations of test case prioritization techniques have been widely proposed in the past~\cite{Rothermel1999,Rothermel1997,Rothermel2001,Elbaum2014,Elbaum2002,Epitropakis2015,Hemmati2015,Jiang2015,Jones2003,Korel2005,Korel2008,Luo2016,Luo2019,Henard2016}.
One of the first empirical studies in test case prioritization techniques was proposed by Rothermel et al.~\cite{Rothermel1999}. Specifically, they compared several total and additional test case prioritization techniques, in addition to random and unordered test case prioritization for applications programmed in C. Do et al. proposed an evaluation of additional and total strategies to test Java programs tested under the JUnit framework~\cite{Do2004}. Luo et al. compared the performance of static and dynamic test case prioritization techniques applied in Java programs~\cite{Luo2016,Luo2019}. A major finding was that static techniques performed best at test-class level whereas dynamic ones performed better at test-method level when considering APFD values. Shin et al.~\cite{Shin2019} empirically evaluated how mutation testing criteria (traditional and diversity-aware) could be employed as test quality metric to prioritize test cases. Their findings included that there was no single dominant technique. Similar to this work, Henard et al. proposed a comparison between black-box and white-box test case prioritization techniques~\cite{Henard2016}. They found little difference between white-box and black-box techniques. The main difference between prior empirical evaluations and the study we propose is related to the type of systems being targeted. 
While most of the empirical works for test case prioritization focus on testing languages like C or Java, in this paper we propose and evaluate techniques for test case prioritization designed for Simulink models, the main models used to simulate CPSs. The main difference between programming languages like Java or C  with respect to Simulink is the continuous-time semantics of Simulink with respect to discrete semantics of such languages. This leverages opportunities to propose new test quality metrics, such as those proposed at the signal-level and considered as ``anti-patterns''. While there are other modeling/simulation languages with continuous-time semantics (e.g., Labview or BCVTB), to the best of our knowledge, this is the first work that performs an empirical evaluation of classical test case prioritization techniques in the context of modeling/simulation languages.

\section{Conclusion}
\label{sec:Conclusion}
This paper conducted, what, to the best of our knowledge is, the first empirical study of test case prioritization techniques for the context of Simulink models. Specifically, we analyzed a total of 11 black-box and white-box techniques for test case prioritization adapted to Simulink models. Our key finding were interesting. On the one hand, we found that, unlike for the context of test case selection~\cite{Arrieta2019a}, for the context of test case prioritization white-box techniques are slightly more competitive than black-box techniques. Another finding was that, in general, total greedy-based test case prioritization techniques were more effective than additional greedy-based test case prioritization techniques. On the other hand, as expected and similar to the context of test case selection~\cite{Arrieta2019a}, we found that black-box techniques that rely on anti-patterns to measure the quality of test cases are more effective than similarity-based black-box test case prioritization techniques. Overall, results may differ from a system to another, although most techniques provide consistently high APFD values, suggesting that they are appropriate for test case prioritization. Our recommendations in Section \ref{sec:CR} provides more insights and specific recommendations based on the setup of practitioners (e.g., availability of access to white-box metrics, models size).  

\textbf{Data Availability:} We make all the scripts, model and results available: \url{https://doi.org/10.6084/m9.figshare.28777829}



\textbf{\subsection*{Acknowledgments} }
 The author is part of the Software and Systems Engineering research group of Mondragon Unibertsitatea (IT1519-22), supported by the Department of Education, Universities and Research of the Basque Country. The author would like to thank to Prof. Gregg Rothermel for comments and feedback in a prior version of the manuscript.

\bibliographystyle{plain}
\bibliography{bibliography}

\end{document}